\documentclass[fleqn,usenatbib]{mnras}

\usepackage{newtxtext,newtxmath}
\usepackage[T1]{fontenc}

\DeclareRobustCommand{\VAN}[3]{#2}
\let\VANthebibliography\thebibliography
\def\thebibliography{\DeclareRobustCommand{\VAN}[3]{##3}\VANthebibliography}

\normalsize
\normalsize
\normalsize

\usepackage{graphicx}	% Including figure files
\usepackage{amsmath}	% Advanced maths commands
\usepackage{fix-cm}
\title[LBG Cosmology]{Impact of redshift distribution uncertainties on Lyman-break galaxy cosmological parameter inference}

\author[F. Petri et al.]{Francesco Petri,$^{1}$\thanks{E-mail: francesco.petri14@imperial.ac.uk}
Boris Leistedt$^{1}$,
Daniel J. Mortlock$^{1, 2}$,
Joel Leja$^{3, 4, 5}$,
Stephen Thorp$^{6}$,
Justin Alsing$^{6}$,
\newauthor{Hiranya V. Peiris$^{7, 6}$,
Sinan Deger$^{7}$}\\
\\
$^{1}$Department of Physics, Imperial College London, Blackett Laboratory, Prince Consort Road, London SW7 2AZ, UK\\
$^{2}$Department of Mathematics, Imperial College London, London, SW7 2AZ, UK\\
$^{3}$ Department of Astronomy \& Astrophysics, The Pennsylvania State University, University Park, PA 16802, USA \\
$^{4}$ Institute for Computational \& Data Sciences, The Pennsylvania State University, University Park, PA 16802, USA \\
$^{5}$ Institute for Gravitation \& the Cosmos, The Pennsylvania State University, University Park, PA 16802, USA\\
$^{6}$ The Oskar Klein Centre, Department of Physics, Stockholm University, AlbaNova University Centre, SE 106 91 Stockholm, Sweden \\
$^{7}$ Institute of Astronomy and Kavli Institute for Cosmology, University of Cambridge, Madingley Road, Cambridge, CB3 0HA, UK
}
\date{Accepted XXX. Received YYY; in original form ZZZ}
\pubyear{2025}

\begin{document}
\label{firstpage}
\pagerange{\pageref{firstpage}--\pageref{lastpage}}
\maketitle

\begin{abstract}
 A significant number of Lyman-break galaxies (LBGs) with redshifts $3 \lesssim z \lesssim 5$ are expected to be observed by the upcoming Vera C.\ Rubin Observatory Legacy Survey of Space and Time (LSST). This will enable us to probe the universe at higher redshifts than is currently possible with cosmological galaxy clustering and weak lensing surveys. However, accurate inference of cosmological parameters requires precise knowledge of the redshift distributions of selected galaxies, where the number of faint objects expected from LSST alone will make spectroscopic based methods of determining these distributions extremely challenging. To overcome this difficulty, it may be possible to leverage the information in the large volume of photometric data alone to precisely infer these distributions. This could be facilitated using forward models, where in this paper we use stellar population synthesis (SPS) to estimate uncertainties on LBG redshift distributions for a 10 year LSST (LSSTY10) survey. We characterise some of the modelling uncertainties inherent to SPS by introducing a flexible parameterisation of the galaxy population prior, informed by observations of the galaxy stellar mass function (GSMF) and cosmic star formation density (CSFRD). These uncertainties are subsequently marginalised over and propagated to cosmological constraints in a Fisher forecast. Assuming a known dust attenuation model for LBGs, we forecast constraints on the $\sigma_{8}$ parameter comparable to {\em Planck} cosmic microwave background (CMB) constraints. 
 
\end{abstract}


%%%%%%%%%%%%%%%%%%%%%%%%%%%%%%%%%%%%%%%%%%%%%%%%%%

%%%%%%%%%%%%%%%%% BODY OF PAPER %%%%%%%%%%%%%%%%%%

\section{Introduction}
\label{sec:intro}
With a new generation of surveys such as the Vera C. Rubin Observatory Legacy Survey of Space and Time (LSST) \citep{lsst_book, ivezic19} and \textit{Euclid} \citep{mellier24},
we will have access to larger numbers of galaxies than ever before. The Rubin Observatory is expected to observe several billion galaxies over an area up to 18,000 $\mathrm{deg}^{2}$, with number densities potentially as high $\sim 30$ $\mathrm{arcmin}^{-2}$ \citep{chang13, desc18, ivezic19}. Harnessing this statistical power for constraining cosmology is one of the key science goals for LSST, which will probe the dark energy and dark matter distribution in the universe up to redshifts $z \sim$ 2--3 \citep{ivezic19, desc18}.

With the increased sensitivity of these next generation instruments, it is expected that we will be able to observe significant numbers of higher redshift galaxies at $z \sim$ 3--5 \citep{wilson19}. The most straightforward high redshift galaxy to observe is via the dropout technique, which selects for galaxies with bright, strong Lyman breaks (Lyman-break galaxies; LBGs) \citep{g1990, steidel96, giava2002, shap11}. These galaxies are typically young, star forming galaxies containing containing many massive, highly luminous O and B type stars which dominate the galaxy spectral energy distribution (SED) \citep{giava2002, shap11}. Absorption of light from these galaxies by intervening neutral hydrogen present in stellar atmospheres, the interstellar medium (ISM), and the intergalactic medium (IGM), produces the Lyman break. This is readily targeted by photometric surveys, where for example, Subaru Hyper Suprime-Cam (HSC) \citep{goldrush4} identified $\sim\!4\times10^6$ LBG candidates in the redshift range $2 \lesssim z \lesssim7$.

While LBGs have been studied for a long time \citep{steidel96}, more recent work \citep{yu2018, schmittfull2018, wilson19, miya22} has motivated the use of LBGs for cosmological applications, in time for next generation large scale surveys such as LSST. By measuring the angular clustering of LBGs, we can, via cross-correlations with the CMB lensing signal, probe the matter distribution of the universe before the dark energy-dominated era \citep{wilson19}. As such, by using LBGs, we will have access to the universe on scales and redshifts mostly out of reach of fiducial weak lensing surveys, providing increased sensitivity to early universe physics. This has been proposed to allow for more stringent tests on General Relativity (GR) and for constraining primordial non-Gaussianity (PNG) \citep{wilson19}. With the target redshift range of $z\sim$ 3--5, LBGs could provide (independently from current galaxy surveys) a measurement of the parameter $S_{8} \equiv \sigma_{8}(\Omega_{\mathrm{m}}/0.3)^{0.5}$ (here $\Omega_{\mathrm{m}}$ is the matter density and $\sigma_{8}$ is the root mean square variation of matter in spheres of radius 8 $\mathrm{Mpc}/h$, with $h$ the dimensionless Hubble parameter). This could be exploited to inform the current `tension' between weak lensing \citep[e.g.,][]{des3, hsc3, hsc32} and cosmic microwave background (CMB) \citep{planck18} measurements of $S_{8}$. LBG clustering may also be exploited to constrain the neutrino mass \citep{yu2018}. 

However, such constraints can only be obtained from angular clustering measurements with accurate knowledge of $N(z)$, the redshift distribution of the galaxies in the sample. This encodes the distance information which links the projected clustering of the galaxies on the sky to the full spatial distribution. Incorrect determination of $N(z)$ is known to bias constraints on cosmological parameters \citep{nzbias1, nzbias2}, so it is vital to estimate the redshift distribution accurately.

There are a number of ways to estimate $N(z)$ for a photometric galaxy survey. Frequently this involves spectroscopy, leveraging accurate spectroscopic measurements to calibrate the less reliable photometric data. One way this can be done is via cross-correlation \citep[e.g.,][]{newman2008,hilde2017,davis18,gatti2021}, where the target photometric galaxy population is cross-correlated with another population with known spectroscopic redshifts. Where  spectroscopic redshifts are available for some representative subset of the photometry, direct calibration \citep[e.g.,][]{lima2008, kidsred} is also possible. These methods are limited by the availability of spectroscopic redshifts, which for the numbers and depths probed by surveys such as LSST will be prohibitively time consuming, particularly for fainter high-redshift galaxies such as LBGs. Also these methods can introduce biases caused by spectroscopic selection effects, which can be difficult to model. As a result, these can lead to poor coverage of faint, higher redshift galaxies with weaker spectral features \citep{newman2022}.

Therefore, it is necessary to consider methods for inferring $N(z)$ without the use of spectroscopy. One approach, called template methods (see \citealp{salvato2019, newman2022} for reviews), assume galaxies belong to certain `types', with associated SED templates. These templates are made up of libraries of observed or theoretical galaxy SEDs, which when compared to observed photometry can be used to infer galaxy redshift distributions. However, by virtue of being limited to a finite set of SEDs, one is restricted to a discrete grid of the SED parameter space sampled by the templates. This results in areas of the SED prior volume being missed, making it difficult to model galaxy types not explicitly included in the templates.

To overcome this limitation, Stellar Population Synthesis (SPS) (see \citealt{spsrev} for a review) can be used to simulate large numbers of galaxy SEDs in a more continuous manner. This requires making assumptions about the target population of galaxies, i.e., priors on physical parameters such as redshift, dust content, metallicity etc., which are passed to the SPS model to generate SEDs. Recent advances in machine learning (ML) methods make it easier to simulate a large quantities of synthetic galaxy SEDs \citep{speculator}, which can facilitate forward modelling galaxy redshift distributions as shown in recent work by \citet{alsing2023}.

Redshift distribution modelling for LBGs has been attempted previously for the Canada-France-Hawaii Telescope Legacy Survey (CFHTLS) \citep{cars09}. This highlighted how different template-based methods and SPS simulations yield different redshift distributions, with differing amounts of low redshift interlopers. These are low redshift galaxies that have features that mimic the Lyman break and contaminate the sample. The fraction of interlopers that make it into the sample will be extremely important to quantify, as these can bias constraints on cosmological parameters, as shown by \citet{wilson19}. However, previous forecasts on cosmological parameters using LBGs from LSST have either assumed known or simplistic handling of LBG redshift distribution uncertainties \citep{yu2018, schmittfull2018, wilson19}. Quantifying the impact of low redshift interlopers on LBG redshift distributions presents a challenge, as future large spectroscopic surveys may not be able to provide secure redshifts over a large enough fraction of LBGs between $z =$ 3--5. The upgrade to the Dark Energy Spectroscopic Instrument (DESI), known as DESI-II, may provide spectra for up to 50--80\% of selected LBG targets at $z \sim 3$ for a photometric survey such as LSST \citep{ruhlmann24}. Forward modelling could be the only way to exploit photometry for the remaining galaxies, in particular for fainter ones at higher redshifts of $z \sim$ 4--5. This will require leveraging information of the physical properties of the galaxy population up to $z \sim$ 5--6.

In this work we present a method to predict LBG redshift distributions by building a forward galaxy population model inspired by the work done in \citet{cars09} and \citet{alsing2023}. Importantly, we also model the uncertainty in the distributions, by introducing flexibility in the galaxy population model. We achieve this by fitting Gaussian processes to observed galaxy stellar mass functions (GSMFs) and the cosmic star formation rate density (CSFRD). This allows us to sample different realisations of the galaxy population prior, calibrated by observational measurements. We can then define a redshift distribution prior, which is subsequently marginalised over to produce a Fisher forecast on cosmological parameters for an LSSTY10 style survey. After reviewing the required galaxy clustering theory in Section~\ref{sec:background}, the forward model used is described in Section~\ref{sec:forwardmodel}, with the associated error propagation and forecast cosmological constraints presented in Section~\ref{sec:comsoforecast}. The limitations and potential extensions of our approach are discussed in Section~\ref{sec:discussions} and our conclusions presented in Section~\ref{sec:conclusions}.

\section{Background}
\label{sec:background}

Galaxies are typically used to probe the underlying cosmological matter distribution of the universe by measuring their clustering combined with gravitational weak lensing lensing measurements. We provide a brief overview of the statistical description of clustering in Section \ref{sec:powerspec}. It has been proposed that for LBGs between $3 \lesssim z \lesssim 5$, cross-correlations with the CMB lensing signal can assist in constraining the galaxy bias \citep{yu2018, schmittfull2018, wilson19}. We introduce this in Section \ref{sec:cmblens}.

\vfill%%%%%%%%%%%%%%%%%%%%%%%%%%%%%%%%%%%%%

\subsection{Galaxy clustering}
\label{sec:powerspec} 

In order to probe the matter distribution of the universe using LBGs, we can measure their clustering. The matter density in the universe at redshift $z$, at a position in in 3D space $\mathbf{x}$ is typically expressed as an over-density:
\begin{equation}
    \delta(\mathbf{x}, z) = \frac{\rho(\mathbf{x}, z) - \bar{\rho}}{\bar{\rho}}.
\end{equation}
The quantity $\rho(\mathbf{x})$ is the matter density at $\mathbf{x}$ relative to the mean density $\Bar{\rho}$. The clustering of matter between pairs of points at a given $z$ can be quantified using the 2-point correlation function
\begin{equation}
    \xi(r, z) = \langle \delta(\mathbf{x}, z) \delta(\mathbf{x} + \mathbf{r}, z) \rangle ,
    \label{eq:corfunc}
\end{equation}
where $r$ is the distance between the two points \citep{bern2002}. This is typically expressed in Fourier space as the matter power spectrum $P_{\mathrm{m}}$, defined as the Fourier transform of equation \ref{eq:corfunc}:
\begin{equation}
    P_{\mathrm{m}}(k, z) = \int \xi(r, z) \exp^{-i\mathbf{k}\cdot\mathbf{r}}d^{3}\mathbf{r},
\end{equation}
where $k$ is the Fourier mode. We measure the positions of galaxies however, so the matter distribution must be linked to the observed galaxy clustering. On large scales, this is can be achieved by assuming a linear galaxy bias model described by a parameter $b_{d}$ such that
\begin{equation}
    P_{d}(k, z) = b_{d}^{2}P_{\mathrm{m}}(k, z),
    \label{eq:biasdef}
\end{equation}
where $P_{d}$ is the galaxy power spectrum for a given LBG dropout $d$ \citep{biasrev, baumann2022}. By measuring the positions of galaxies, we can relate the measured clustering to the underlying matter distribution. This is a simple model, and in reality parameter $b_{d}$ may also be a function redshift and scale such that $b_{d} = b_{d}(k, z)$, which can vary for different galaxy populations \citep{baumann2022}. For an overview of different bias models see \citet{biasrev, nicola24}.

However, in the absence of precise redshifts, e.g. from spectroscopic measurements, we can instead leverage the statistical information of the galaxies via the redshift distribution, $N(z)$. Therefore, for a given LBG dropout, we measure the fluctuations of the projected galaxy density on the sky \citep{dodel}, defined as
\begin{equation}
    \Delta_{d}(\hat{\mathbf{n}}) = \int W_{d}(\chi) \delta(\hat{\mathbf{n}} \chi, \eta(\chi))d\chi,
    \label{eq:projmat}
\end{equation}
where $\hat{\mathbf{n}}$ is a two dimensional vector pointing to a direction on the sky, $\chi$ is the co-moving distance, $\eta$ is conformal time and $W_{d}(\chi)$ is the window in which we probe the matter density. This is related to the redshift distribution of a given galaxy population $N_{d}(z)$ as:
\begin{equation}
    W_{d}(\chi) = b_{d} \frac{1}{\bar{N}_{d}} \frac{dN_{d}(z)}{d\chi},
    \label{eq:galkernel}
\end{equation}
where $\bar{N}_{d}$ is the total number of galaxies. We assume different LBG populations will have different galaxy linear bias parameters $b_{d}$, as shown in \citet{wilson19}. 

Clustering between pairs of points on the sky is quantified by the galaxy angular power spectrum $C^{d}_{\ell}$, which can be interpreted as a projection of the 3D power spectrum onto the sky. In the \citet{limber53} approximation this is
\begin{equation}
        C^{d}_{\ell} = \int \frac{d\chi}{\chi^{2}}(W_{d}(\chi))^{2}P_{\mathrm{m}}\bigg(k=\frac{\ell+1/2}{\chi},\eta(\chi)\bigg),
        \label{eq:angpowersimple}
\end{equation}
where $\ell$ is the spherical harmonic index \citep{dodel}.

Therefore, if we would like to extract information on the matter distribution of the universe from galaxy clustering observed on the sky, we must also know $N_{d}(z)$. The redshift distributions encode the critical distance information that relates the projected signal back to cosmological parameters via $P_{\mathrm{m}}$. This means any uncertainty in $N_{d}(z)$ will affect constraints on cosmological parameters.

\subsection{CMB lensing cross-correlations}
\label{sec:cmblens} 

In order to relate observed clustering of galaxies to the underlying matter distribution of the universe, we need to know $b_{d}$ (Equation~\ref{eq:biasdef}). Therefore, cosmological analyses using galaxy clustering requires extra information to quantify correlations between galaxies and matter. For LBGs, CMB lensing cross correlations are a promising probe for this \citep{schmittfull2018, yu2018, wilson19}. 

Matter in the universe distributed between us 
and the surface of last scattering at redshift $z_{*}\sim 1100$, can deflect the trajectory CMB photons as they travel through the universe. So if we cross-correlate the lensing signal with galaxy positions, we probe the relationship between galaxy clustering and the matter density field, so we can constrain $b_{d}$. The matter density probed by CMB photons is quantified by the convergence
\begin{equation}
    \kappa(\hat{\mathbf{n}}) = \int_{0}^{\chi_{*}}W_{\kappa}(\chi)\delta(\chi\hat{\mathbf{n}}, \eta(\chi))d\chi,
\end{equation}
where $\chi_{*}$ is the proper distance to the surface of last scattering and $W_{\kappa}(\chi)$ is the CMB lensing kernel.  This is given by
\begin{equation}
    W_{\kappa}(\chi) = \frac{3 \Omega_{\mathrm{m}}}{2c^{2}}H^{2}_{0}(1+z)\chi(z)\frac{\chi_{*}-\chi(z)}{\chi_{*}},
    \label{eq:cmbkernel}
\end{equation}
where $H(z)$ is the Hubble parameter, with $H_{0} = H(0)$ evaluated at the present day, and $c$ is the speed of light \citep{cmbrev, lensingrev}. Equation \ref{eq:cmbkernel} peaks and decays slowly beyond $z \sim 1$, meaning the CMB probes matter more strongly at redshifts similar to LBGs predicted to be observed by LSST. This makes these two probes ideal for cross-correlations, as illustrated in Figure 5 of \citet{wilson19}. 

For two cosmological probes $x$ and $y$, such as a dropout $d$ or CMB lensing convergence $\kappa$,
Equation~\ref{eq:angpowersimple} is modified such we measure the angular power cross- and auto-spectra $C^{xy}_{\ell}$ as:
\begin{equation}
        C^{xy}_{\ell} = \int \frac{d\chi}{\chi^{2}}W_{x}(\chi)W_{y}(\chi)P_{\mathrm{m}}\bigg(k=\frac{\ell+1/2}{\chi},\eta(\chi)\bigg).
        \label{eq:angpower}
\end{equation}
The kernels $W_{x}(\chi)$ and $W_{y}(\chi)$ can take the form of either Equation~\ref{eq:galkernel} or \ref{eq:cmbkernel}, as seen in \citet{modi2017}. This way we also include cross-correlations between galaxy bins, which has been shown to improve cosmological constraints \citep{schaan2020}.

\section{Forward-modelling redshift distributions}
\label{sec:forwardmodel}

\begin{table*}
 \caption{The full list of parameters for the SPS model used as part of the forward model in this work. These $N_{\mathrm{SPS}}$=16 parameters, denoted by the vector $\boldsymbol{\psi} = (\psi_{1}, \psi_{2},...,\psi_{N_{\mathrm{SPS}}})^{T}$ are sampled from the priors detailed in \S\ref{sec:popmodel}, within the bounds given in the final column in this table.} 
 \label{tab:spsparams}
 \begin{tabular}{lccc}
  \hline
  \hline
  SPS Parameter& Symbol & Bounds \\
  \hline
   Redshift&$z$&(0, 7]\\
    Stellar Metallicity&$\mathrm{log}_{10}(\mathrm{Z}/\mathrm{Z}_{\sun})$&[-2.5, 0.5] \\
   Normalisation of Birth-Cloud Dust Attenuation Curve &$\tau_{1}$ &[0,4]\\
    Normalisation of Diffuse Dust Attenuation Curve &$\tau_{2}$&[0, 4]\\
   Offset to \citet{calzetti2000} Attenuation Curve &$\delta$ &[-2.2, 0.4]\\
   %Fluctuations in IGM optical depth around \citet{madau1995} &$f_{\mathrm{IGM}}$ &[0, 2] \\
   Gas Ionisation& $\mathrm{log}_{10}U$ &[-4, -1]\\
   Gas Phase Metallicity &$\mathrm{log}_{10}(\mathrm{Z}_{\mathrm{gas}}/\mathrm{Z}_{\sun})$ &[-2, 0.5]\\
   AGN Contribution to SED & $\mathrm{log}_{10}(f_{\mathrm{agn}})$ &[-5, 1]\\
   Optical depth of AGN Torus &$\tau_{\mathrm{agn}}$ &[5, 150]\\
   SFR ratios &$\mathbf{x}_{\mathrm{sfr}}$ &[-5, 5]\\
  Log Stellar Mass (in Solar Masses) &$\mathcal{M} = \mathrm{log}_{10}(\mathrm{M}/\mathrm{M}_{\sun})$ &[7,13]\\
  \hline
  \hline
 \end{tabular}
\end{table*}

Our approach to estimating the redshift distribution of a galaxy sample is to construct a forward model of the relevant galaxy population, giving us explicit control over the physical assumptions and approximations we make.  This will also allow us to estimate the uncertainty in observed LBG redshift distributions. The model is split into two main parts: the population model; and the SPS model. The population model is effectively the prior, which describes the population of LBGs in terms of distributions of their physical quantities, (e.g., redshift, metallicity or mass). The SPS model provides the framework for generating galaxy SEDs, and hence photometry, given the physical quantities as inputs. By making draws of physical parameters from the prior and passing them to the SPS model, we can generate photometric data for realistic galaxies, which after applying noise and selection cuts, gives us a set of redshift distributions. An example of forward modelling redshift distributions in this manner using SPS can be found in \citet{alsing2023}.

By keeping the priors fixed, samples of SPS parameters, denoted by vector $\boldsymbol{\psi} = (\psi_{1}, \psi_{2},...,\psi_{N_{\mathrm{SPS}}})$ can be drawn, where $N_{\mathrm{SPS}}$ is the number of SPS parameters needed to describe a single galaxy. When passed to the SPS model to generate photometry, this will yield a single set of LBG redshift distributions. For this forecast, we want to propagate uncertainties from the SPS and galaxy population model to the cosmological constraints, in order to characterise the uncertainty in the modelling of the LBG redshift distributions. So instead, we vary the galaxy population priors, while choosing a flexible SPS model to give multiple realisations of possible redshift distributions, which can be subsequently marginalised over for the cosmological analysis.

We now detail the components of the LBG redshift distribution forward model: the SPS model (Section~\ref{sec:spsmodel}); the galaxy population model (Section~\ref{sec:popmodel}); and the photometric noise model and LBG selection cuts (Sections~\ref{sec:photonoise} and \ref{sec:selec}, respectively). We also describe the SPS emulation used in Section \ref{sec:emulation}. The resulting redshift distributions from the forward model are presented in Section \ref{sec:nzs}.

\subsection{SPS model}
\label{sec:spsmodel} 

The SPS model provides the means for simulating galaxy SEDs, which can then be used to generate LBG redshift distributions. It uses a set of SPS parameters, $\boldsymbol{\psi}$, to generate a single galaxy SED. In essence, the total galaxy SED is a superposition of stellar spectra, with several other physical effects also included. See \citet{spsrev} for a detailed review.

We implement our model using Python FSPS \citep{johnson2024, conroy2009, conroy2010}, following the choices used in \textsc{Prospector-}$\beta$ \citep{wang2023}, with the inclusion of the gas ionisation parameter as in \citet{popcosmos}. We adopt the parameterisation used in \citet{popcosmos}, where we parametrise our SPS model by the $N_{\mathrm{SPS}}=16$ parameters listed in Table \ref{tab:spsparams}. We have assumed MIST isochrones \citep{mist1, mist2, mist3, mist4, mist5}, MILES stellar spectral libraries \citep{miles2, miles1}, and a Chabrier initial mass function (IMF) \citep{chabrier2003}. Nebular emission is calculated using CLOUDY \citep{ferland2013, cloudy1}. This is important to model for star forming galaxies such as LBGs, as UV-optical light from young stars in these galaxies drives nebular emission. 

We assume a \citet{drainelie2007} dust emission model as implemented in FSPS, but as this work focuses on rest frame UV photometry, choice of dust emission model should have a negligible impact on our results. For modelling emission from active galactic nuclei (AGN), we follow \citet{leja2018}, which uses the FSPS implementation of the CLUMPY AGN templates from \citet{clumpy1, clumpy2}. This allows for two free parameters: $f_{\mathrm{AGN}}$ and $\tau_{\mathrm{AGN}}$, which govern the fractional bolometric luminosity and the dust torus optical depth of the AGN respectively. 

For the current implementation of our SPS model, we evaluate the luminosity distance and the age of the universe using a WMAP9 \citep{bennett2013} cosmology, guided by the default cosmology used by FSPS. The possible uncertainty in simulated redshift distributions introduced by a mismatched cosmology (e.g., between WMAP9 and the most recent \textit{Planck} results \citep{planck18}) is negligible compared to the uncertainty we introduce by varying the galaxy population model in Section~\ref{sec:popmodel}. Therefore this choice of cosmology will be sufficient for the purposes of this work. 

We have detailed our SPS modelling choices in the sections that follow; the limitations are discussed in Section~\ref{sec:spslimitations}.

\subsubsection{Star formation history}
\label{sfh}
The star formation history (SFH) describes the star formation rate (SFR) of a galaxy as a function of time. We have chosen the continuity model \citep{sfhnonparam} to describe the SFH of our simulated galaxies, which divides the SFH into bins parametrised by the log-SFR ratio between each bin. So for a given bin $n$, the log SFR ratio between bins $n$ and $n+1$ is given as $x=\mathrm{log}_{10}(\mathrm{SFR}_{n}/\mathrm{SFR}_{n+1})$. Following \cite{sfhnonparam}, we adopt a Student's t distribution as a prior on these ratios, given by
\begin{equation}
    t(x, \mu, \sigma, \nu) = \frac{\Gamma(\frac{\nu + 1}{2})}{\sqrt{\nu \pi}\Gamma(\frac{\nu}{2})} \bigg( 1 + \frac{(x/\sigma)^{2}}{\nu}  \bigg)^{-\frac{\nu+1}{2}},
\end{equation}
where $\nu$ is the degrees of freedom parameter fixed at $\nu=2$, $\sigma$ characterises the width of the distribution, $\mu$ is the mean, and $\Gamma(\cdot)$ is the Gamma function. We use a total of seven bins to describe the SFH, the two most recent bins are defined as [0, 30] and [30, 100] $\mathrm{Myr}$ in lookback time. The oldest bin is scaled by the age of the universe $t_{\mathrm{age}}(z)$ such that it is given by $[0.85, 1] \,t_{\mathrm{age}}(z)$, with the remaining four bins being spaced logarithmically in time. The use of the continuity model allows the forward model to generate galaxies with a wide range of different SFHs, much more than would be possible with more restrictive `parametric' forms of the SFH \citep{sfhparam, sfhnonparam}. In this work we will vary $\mu$ and fix $\sigma = 0.3$.

\subsubsection{Dust attenuation model}
\label{dustatt}
The dust attenuation model describes how the light emitted by a source galaxy is attenuated as a function of wavelength, $\lambda$, before reaching the observer. This is described by an attenuation curve $\tau_{\lambda}$, which encodes the effect of a variety of different processes, including absorption, the geometry of stars/dust and the scattering of light, both into and out of the line of sight, on the final observed distribution of light from a given galaxy (see \citealp{dustreview} for a review). 

We follow the modelling choices as described in \citet{leja2017}, and assume a two component model as proposed in \citet{charlotfall2000}, where light is attenuated by a birth-cloud and diffuse dust screens. The birth-cloud screen only attenuates stars with ages less than $10$ Myr (the maximum lifetime of a molecular cloud, as given by \citealt{blitzshu1980}).
The optical depth of the birth-cloud screen $\tau_{\lambda, 1}$ is assumed to vary such that
\begin{equation}
    \tau_{\lambda, 1} = \tau_{1}\bigg(\frac{\lambda}{5500 \  \text{\AA}}\bigg)^{-1.0},
\end{equation}
where $\tau_{1}$ is a free parameter controlling the normalisation of the birth-cloud attenuation curve. Also as described in \citet{leja2017}, the optical depth of the diffuse dust screen $\tau_{\lambda, 2}$ is given by \citep{noll2009}
\begin{equation}
    \tau_{\lambda, 2} = \frac{\tau_{2}}{4.05}[k'(\lambda)+D(\lambda)]\bigg(\frac{\lambda}{5500 \  \text{\AA}}\bigg)^{\delta},
\end{equation}
where $k'(\lambda)$ is the \citet{calzetti2000} curve and $D(\lambda)$ is a Lorentzian-like `Drude' profile characterising the UV dust bump \citep{fitz1990, noll2009, leja2017}. The shift from the Calzetti attenuation curve, $\delta$, and the diffuse screen normalisation, $\tau_{2}$, are left as free parameters. The strength of $D(\lambda)$ is tied to $\delta$ via Equation~3 from \citet{kriekconroy2013} as implemented in FSPS.

\subsubsection{Intergalactic medium}

The intergalactic medium (IGM) plays a critical role in the shape of galaxy SEDs as the neutral Hydrogen in the IGM (and to an extent in the ISM) drives the absorption blueward of the Lyman-Break \citep{giava2002, shap11}. We model the absorption of light from an observed galaxy by intervening IGM in the line of sight using the FSPS implementation of the \citet{madau1995} model. This quantifies the absorption of light from Poisson-distributed clouds of neutral hydrogen in the line of sight as a function of wavelength, for a given redshift, averaged over sight-lines. 

\subsection{Galaxy population model}
\label{sec:popmodel}

\begin{table*}
 \caption{A list of all the free parameters of the population model used in the paper. Sampling these parameters, which in the text we refer to as $\boldsymbol{\varphi}$, allows us to generate different realisations of the galaxy population model.} 
 \label{tab:popparams}
 \begin{tabular}{lccc}
  \hline
  \hline
  Population Model Parameter & Symbol & Bounds\\
  \hline
  \textbf{Galaxy Stellar Mass Function (GSMF) Parameters}\\
   GSMF Normalisations & $\boldsymbol{\phi}^{*}$  & -\\
    GSMF Slopes & $\boldsymbol{\alpha}$ & - \\
    Log Characteristic Stellar Masses (in Solar Masses) & $\boldsymbol{\mathcal{M}_{*}}$ & - \\
    \\
    \textbf{Cosmic Star Formation Rate Density (CSFRD) Parameters}\\
    CSFRD evolution & $\boldsymbol{\rho}$ & - \\
    \\
    \textbf{Metallicity and Gas Ionisation}\\
    Population Mean Log AGN Contribution to SED & $\mu_{\mathrm{agn}}$ & [-5, 1] \\
    Population Mean Optical Depth of AGN Dust Torus & $\mu_{\tau}$ & [5, 150] \\
    Population Mean Gas Log Ionisation Parameter & $\mu_{\mathrm{U}}$ & [-4, -1] \\
    Population Mean Log Stellar Metallicity & $\mu_{\mathrm{Z}}$ & [-2.5, 0.5] \\
    Population Mean Log Gas Phase Metallicity & $\mu_{\mathrm{Z}_{\mathrm{gas}}}$ & [-2, 0.5] \\
    Population Standard Deviation of the Log AGN Contribution to SED & $\sigma_{\mathrm{agn}}$ & [0.01, 6.0] \\
    Population Standard Deviation of Optical Depth of AGN Dust Torus & $\sigma_{\tau}$ & [0.01, 145] \\
    Population Standard Deviation of Log Gas Ionisation & $\sigma_{\mathrm{U}}$ & [0.01, 3.0] \\
    Population Standard Deviation of Log Stellar Metallicity & $\sigma_{\mathrm{Z}}$ & [0.01, 3.0] \\
    Population Standard Deviation of Log Gas Phase Metallicity & $\sigma_{\mathrm{Z}_{\mathrm{gas}}}$ & [0.01, 2.5] \\

   \hline
   \hline
 \end{tabular}
\end{table*}

The galaxy population model is  effectively the prior distribution for the parameters $\boldsymbol{\psi}$ introduced above. This covers the entire galaxy population, not just detected LBGs, as it is necessary to model the way LBGs are selected in real surveys. We observe a broad, noisy sample of galaxies in the universe, then perform cuts to (hopefully) select the LBGs. It is hence necessary to also model the low redshift contaminants present in an LBG sample. The population model $\mathcal{P}$ is denoted as
\begin{equation}
    \label{eq:popdef}
    \mathcal{P}(\boldsymbol{\psi}) = \mathcal{P}(z, \mathcal{M}, \boldsymbol{\psi}_{\mathrm{d}}, \mathbf{x}_{\mathrm{sfr}}, U, Z, Z_{\mathrm{gas}}, f_{\mathrm{agn}}, \tau_{\mathrm{agn}}),
\end{equation}
where the inter-stellar medium dust parameters are denoted as $\boldsymbol{\psi}_{\mathrm{d}} = (\tau_{1}, \tau_{2}, \delta )^{T}$.
Ideally, once LSST data is available (combined with observations in other bands such as in the infrared), one would infer this population prior from the photometry. This has been done by \citet{popcosmos} for the Cosmic Evolution Survey (COSMOS; \citealp{scoville2007, weaver2022}) up to $z\sim 3$. However, for the purposes of this work, we need a model calibrated on observations at $z >3$. Extrapolating the 16 dimensional model from \citet{popcosmos} would be infeasible, as a key insight of that model is that it includes information on higher dimensional correlations between SPS parameters, where there currently is little literature on how these would extend to $z\sim 7$.

Instead we adopt a simpler model, factorising Equation~\ref{eq:popdef} as
\begin{multline}
    \label{eq:popmodelfac}
    \mathcal{P}(\boldsymbol{\psi}) = p(z, \mathcal{M}) p(\mathbf{x}_{\mathrm{sfr}}|z, \mathcal{M})p(\boldsymbol{\psi}_{\mathrm{d}}|\mathbf{x}_{\mathrm{sfr}}) \\
    \times p(U)p(Z)p(Z_{\mathrm{gas}})P(f_{\mathrm{agn}})P(\tau_{\mathrm{agn}}),
\end{multline}
where $p(\cdot)$ denotes a probability density function. Factorising the probabilities in this manner allows us to use measurements of the galaxy stellar mass function (GSMF), cosmic star formation rate density (CSFRD), and the dust vs.\ SFR relationship from \textsc{pop-cosmos} \citep{popcosmos} to calibrate the model up to $z\sim7$. Critically, we parametrise the population model in Equation~\ref{eq:popmodelfac} with a set of parameters $\boldsymbol{\varphi} = (\varphi_{1}, \varphi_{2},...,\varphi_{N_{\mathrm{POP}}})$ such that we can draw different forms of the population model by sampling $\boldsymbol{\varphi}$ from a higher level prior $p(\boldsymbol{\varphi})$. By sampling different population models, we can generate different realisations of LBG redshift distributions. How we parametrise the population model is discussed in the following subsections, where a summary is given in Table~\ref{tab:popparams}.

\subsubsection{Redshift-mass prior}
\label{sec:priorredmass}

The prior $p(z, \mathcal{M})$ is calibrated using measurements of the evolving number density of galaxies in the universe, described by the Galaxy Stellar Mass Function (GSMF). This is typically modelled using a \citet{schechter1976} function, in which the galaxy number density per logarithmic mass is given by
\begin{multline}
    \label{eq:shect}
    \Phi(\mathcal{M}; z) = \phi^{*}\mathrm{ln}(10)e^{-10^{(\mathcal{M} - \mathcal{M}_{*})}}10^{(\mathcal{M}-\mathcal{M_{*}})(\alpha+1)},
\end{multline}
where $\phi^{*}$, $\alpha$ and $\mathcal{M}_{*}$ are free parameters which describe: the overall normalisation, the slope of the curve and logarithmic characteristic stellar mass in solar masses, respectively.  

However, these parameters are expected to evolve with redshift, so we need a set of parameters that can describe the evolution of the GSMF over a given redshift grid. For a grid of $w$ redshifts $\mathbf{z} = ( z_{1}, z_{2}, ..., z_{w})^{T}$, we can describe the redshift evolution as a set GSMFs evaluated at different redshifts $\boldsymbol{\Phi}(\mathcal{M}) = (\Phi(\mathcal{M}; z_{1}), \Phi(\mathcal{M}; z_{2}), ... \Phi(\mathcal{M}; z_{w}))^{T}$. This requires a total $3w$ parameters describing the evolution of $\phi^{*}$, $\alpha$ and $\mathcal{M}_{*}$, denoted as $\boldsymbol{\phi}^{*} = (\phi^{*}_{1}, \phi^{*}_{2}, ..., \phi^{*}_{w})^{T}$, $\boldsymbol{\alpha} = ( \alpha_{1}, \alpha_{2}, ..., \alpha_{w})^{T}$ and $\boldsymbol{\mathcal{M}}_{*} = (\mathcal{M}_{*, 1}, \mathcal{M}_{*, 2}, ..., \mathcal{M}_{*, w} )^{T}$ respectively.

\begin{figure}
    \centering
    \hspace{-7mm}
    \includegraphics[width=0.95\linewidth]{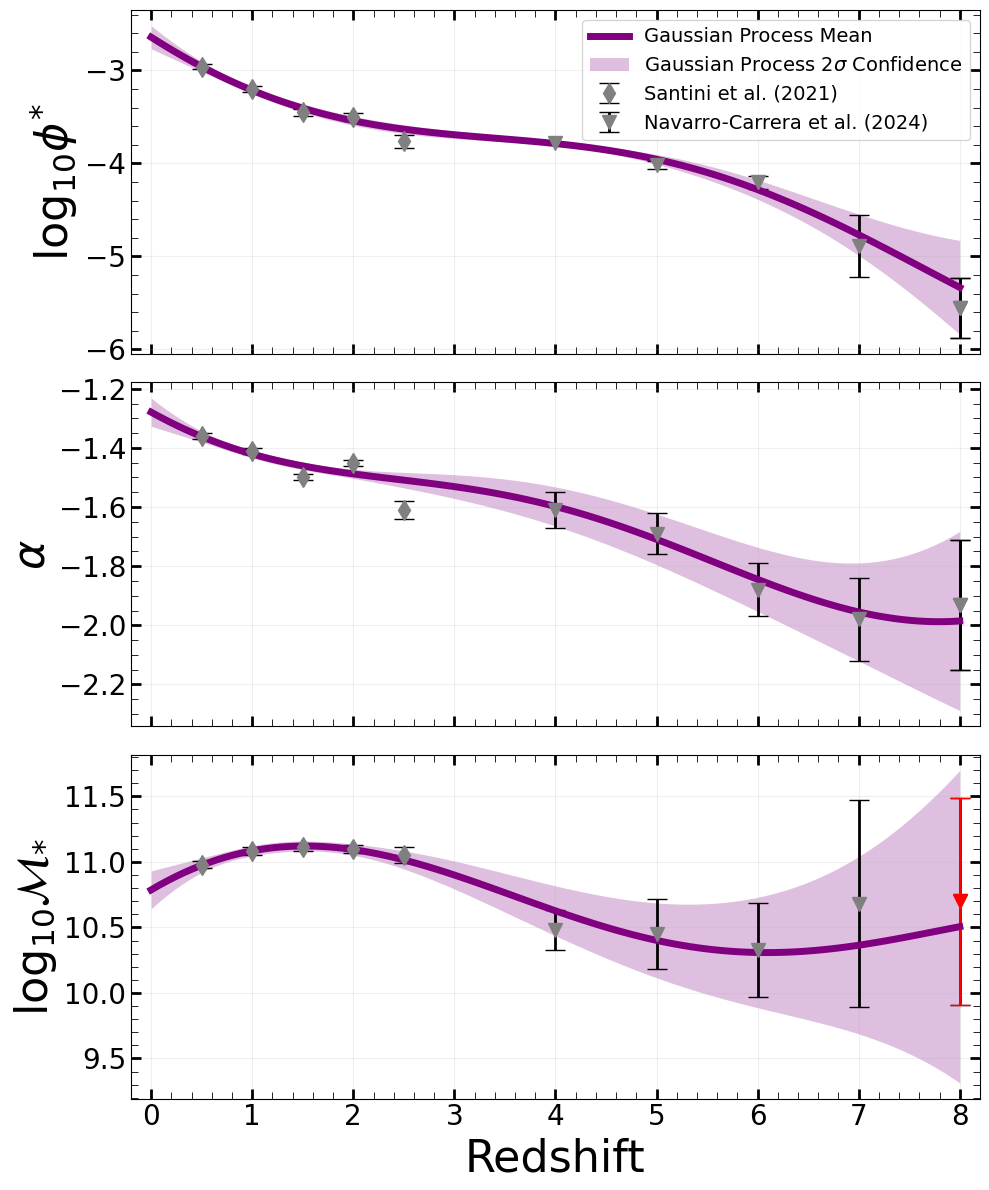}
    \caption{Gaussian process models of the redshift dependence of the  Schechter function parameters using measurements from \citet{santini2022} and \citet{nc2024}. The errors on these data have been made symmetric by taking the average of the upper and lower error bound. Where parameters have been fixed in \citet{nc2024}, highlighted here in red, we have estimated the uncertainties by using the uncertainties quoted for the nearest redshift bin to avoid over-fitting. The solid line shows the Gaussian process mean, with the shaded area showing two standard deviations above and below the mean.}
    \label{fig:gpmassfunc}
\end{figure}

We fit Gaussian processes (GPs) using \textsc{GPyTorch} \citep{gpytorch} to measurements of $\phi^{*}$, $\alpha$ and $\mathcal{M}_{*}$ between $z$ = 0--7 given by \citet{santini2022} and \citet{nc2024}. This is shown in Figure \ref{fig:gpmassfunc}. The advantage of using GPs is that fitting one to observed data defines a distribution of functions that fit this data. This distribution can then be sampled to give different realisations of $\boldsymbol{\phi}^{*}$, $\boldsymbol{\alpha}$ and $\boldsymbol{\mathcal{M}}_{*}$, which as we will show, results in different forms of $p(z, \mathcal{M})$. Similar to previous work \citep{drory2008, peng2010, leja2015, williams2018} this allows us to describe the continuous redshift evolution of the GSMF. However the GP parameterisation gives the added benefit of being able to describe the redshift evolution of the uncertainty in GSMF, which we can propagate to the LBG redshift distributions and marginalise over in a forecast on cosmological parameters.

The data shown in Figure \ref{fig:gpmassfunc} gives us a training set for each Schechter function parameter $\boldsymbol{\phi}_{\mathrm{obs}}^{*}$, $\boldsymbol{\alpha}_{\mathrm{obs}}$ and $\boldsymbol{\mathcal{M}}_{*, \mathrm{obs}}$ evaluated over a grid of redshifts $\mathbf{z}_{\mathrm{obs}}$ each. The GP gives the joint distribution of training data for each parameter evaluated at a grid of $\mathbf{z}_{\mathrm{obs}}$, and the underlying noiseless redshift evolution given by for either $\boldsymbol{\phi}^{*}$, $\boldsymbol{\alpha}$ and $\boldsymbol{\mathcal{M}}_{*}$, over the grid $\mathbf{z}$. Therefore we train a total of three GPs to provide models by which we can draw realisations of redshift evolution as:
\begin{align}
    \boldsymbol{\phi}^{*} &\sim p(\boldsymbol{\phi}^{*} | \boldsymbol{\phi}_{\mathrm{obs}}^{*}, \mathbf{z}, \mathbf{z}_{\mathrm{obs}}, \mathbf{C}_{\mathrm{obs}}^{\phi} ), \label{eq:schectmodel1} \\
    \boldsymbol{\alpha} &\sim p(\boldsymbol{\alpha} | \boldsymbol{\alpha}_{\mathrm{obs}}, \mathbf{z}, \mathbf{z}_{\mathrm{obs}} , \mathbf{C}_{\mathrm{obs}}^{\alpha}),
     \label{eq:schectmodel2} \\
    \boldsymbol{\mathcal{M}}_{*} &\sim p(\boldsymbol{\mathcal{M}}_{*} | \boldsymbol{\mathcal{M}}_{*, \mathrm{obs}}, \mathbf{z}, \mathbf{z}_{\mathrm{obs}}, \mathbf{C}_{\mathrm{obs}}^{\mathcal{M}}),
    \label{eq:schectmodel3}
\end{align}
where $\mathbf{C}_{\mathrm{obs}}^{\phi}$, $\mathbf{C}_{\mathrm{obs}}^{\alpha}$ and $\mathbf{C}_{\mathrm{obs}}^{\mathcal{M}}$ are covariance matrices characterising the observational uncertainty on the data for $\phi^{*}$, $\alpha$ and $\mathcal{M}_{*}$ respectively. These are diagonal, characterising the noise shown by the error bars in Figure \ref{fig:gpmassfunc}. For the choice of kernel $\mathbf{K}$ for fitting the GP, we choose the radial basis function (RBF) kernel, which is given by
\begin{equation}
     \mathbf{K}(\mathbf{z}, \mathbf{z}_{\mathrm{obs}}) = \alpha_{\mathrm{s}} \exp\bigg(\frac{-|\mathbf{z} - \mathbf{z}_{\mathrm{obs}}|^{2}}{2\sigma_{\ell}^{2}}\bigg).
    \label{eq:gpkernel}
\end{equation}
This is parametrised by a kernel length scale $\sigma_{\ell}$ and scale parameter $\alpha_{\mathrm{s}}$. To avoid over-fitting we impose $\sigma_{\ell} > 1.0$ for all the Schechter function parameters as the Schechter  parameter fits used for training the GPs are mostly based on measurements inside redshift bins with widths of $z \sim$0.5--1. In addition, a maximum scale is assumed such that $\sigma_{\ell} < 5.0$ to avoid under-fitting so that the GP interpolates better between redshifts covered by different measurements. The resulting GP fits are also shown in Figure ~\ref{fig:gpmassfunc}.

With a statistical model defined for the redshift evolution of the Schechter function parameters in Equations \ref{eq:schectmodel1}--\ref{eq:schectmodel3}, we can use Equation~\ref{eq:shect}, we can construct a model from which we can draw
\begin{equation}
    \Phi \sim p(\Phi|z, \mathcal{M}, \mathbf{z}, \mathbf{z}_{\mathrm{obs}}, \boldsymbol{\phi}^{*}_{\mathrm{obs}}, \boldsymbol{\alpha}_{\mathrm{obs}}, \boldsymbol{\mathcal{M}}_{\mathrm{*, obs}}, \mathbf{C}_{\mathrm{obs}}^{\phi}, \mathbf{C}_{\mathrm{obs}}^{\alpha}, \mathbf{C}_{\mathrm{obs}}^{\mathcal{M}}).
    \label{eq:drawphi}
\end{equation}
We take $\mathbf{z}$ as a grid of $w=100$ equally spaced redshifts between $z=0$ and $z=7$, and then linearly interpolate to allow sampling of $\Phi$ at an arbitrary redshift $z$. The GPs allow us to interpolate between Schechter fits, giving us an estimate of the  continuous redshift evolution of the galaxy stellar mass function across $0.0 < z \leq 7.0$ . See \citet{leja2020} on how the redshift evolution of the galaxy stellar mass function can be directly inferred from observations between $0 < z \leq 3$. 

\begin{figure}
    \centering
    \includegraphics[width=0.95\linewidth]{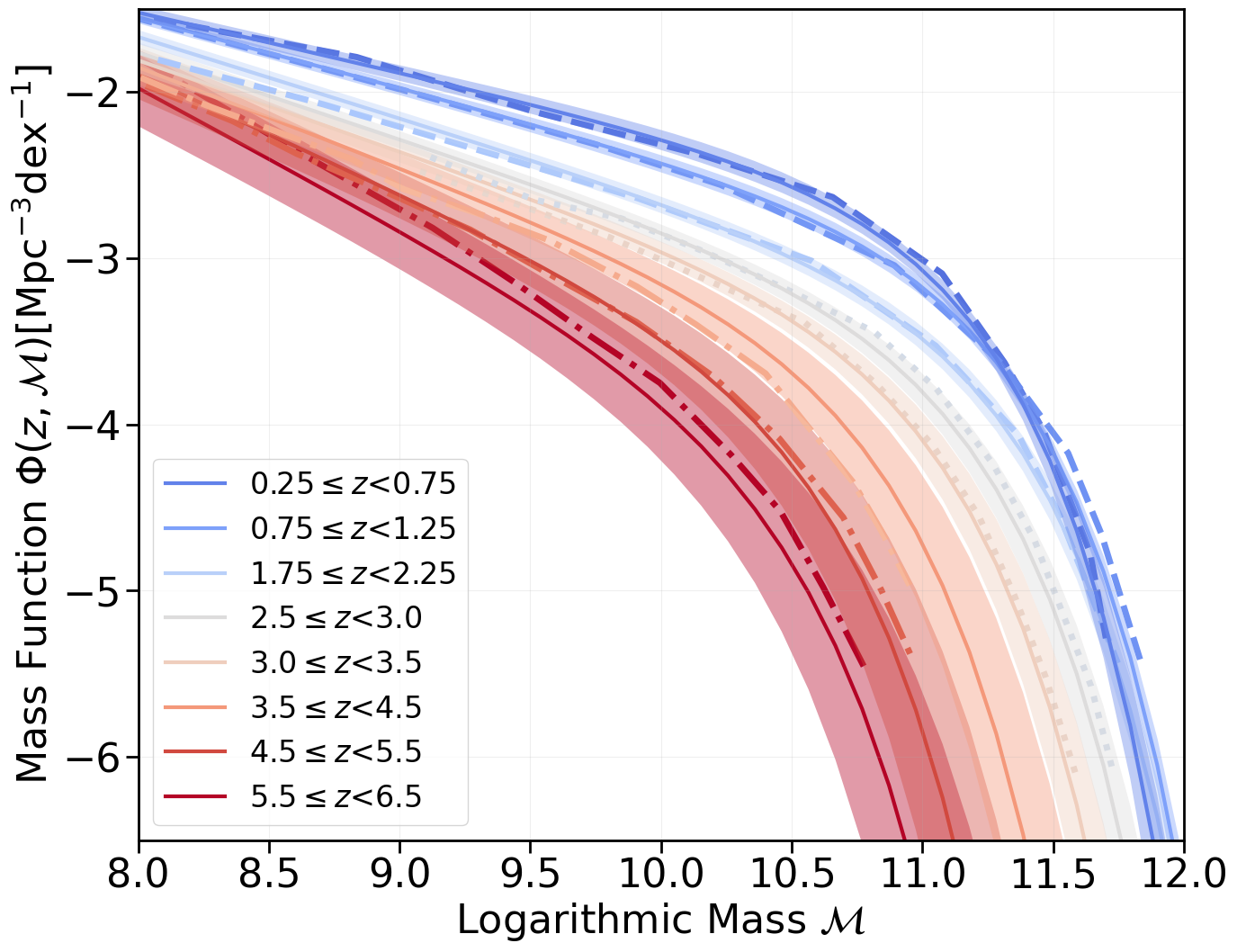}
    \caption{Galaxy stellar mass functions sampled by the Gaussian process model inside a number of redshift bins. The solid lines show the mean, with shaded regions representing the 16-84th percentiles. These are compared with Schechter function fits in \citet{santini2022} (dashed lines), \citet{weaver2023} (dotted lines) and \citet{nc2024} (dashed-dotted lines).}
    \label{fig:massfunclit}
\end{figure}

To check the GP model reproduces the observed mass functions, we compare draws from the distribution in Equation~\ref{eq:drawphi} to other authors' Schechter function fits in Figure~\ref{fig:massfunclit}. These are largely in agreement, with the discrepancies seen being a result of the limitation of modelling the three Schechter function parameters as independent GPs, as the interpolation between the data points in Figure~\ref{fig:gpmassfunc} for a given parameter does not include information in the variation of the other two. However, for our purposes this is sufficient: we want to estimate the uncertainty in mass function measurements to propagate across to redshift distributions and subsequently to cosmological parameters. Figure~\ref{fig:massfunclit} shows that the GP model gives larger uncertainties at higher redshifts, reflecting our lack of empirical knowledge about the galaxy mass function in the early universe. That said, the uncertainty as quantified here may be an overestimate, as we cannot properly combine different measurements with overlapping redshift ranges due to the aforementioned limitation of utilising independent GPs for each parameter. Including more available mass function measurements would be a promising extension of this work.

With a model for drawing samples of $\Phi$ at arbitrary $z$ and $\mathcal{M}$ established in Equation~\ref{eq:drawphi}, we must now normalise $\Phi$ in order to compute $p(z, \mathcal{M})$. This requires calculating the total number of galaxies in the prior volume given by a given realisation of $\Phi$. As we can see from Equation~\ref{eq:shect}, $\Phi \rightarrow \infty$ as $\mathcal{M} \rightarrow - \infty$, for $\alpha < -1$. So if we want to calculate the total number of galaxies, we must set some minimum logarithmic mass (in solar masses) $\mathcal{M}_{\mathrm{lim}}(z)$ to ensure the normalisation factor is finite. We can calculate this as:
\begin{equation}
N_{\mathrm{gal}}=\int_{0}^{\infty}\int_{\mathcal{M}_{\mathrm{lim}}(z)}^{\infty}\Phi(\mathcal{M}, z)\frac{dV_{\mathrm{co}}}{dz}d\mathcal{M}dz,
    \label{eq:numgal}
\end{equation}
where $V_{\mathrm{co}}$ is the co-moving volume. An appropriate choice of $\mathcal{M}_{\mathrm{lim}}(z)$ will be discussed in Section \ref{sec:nzs}. Using Equation~\ref{eq:numgal} we can calculate the redshift-logarithmic mass prior as:
\begin{equation}
    p(z, \mathcal{M}) = \frac{\Phi(\mathcal{M}, z)}{N_{\mathrm{gal}}},
    \label{eq:zmpriordef}
\end{equation}
first shown equation \ref{eq:popmodelfac}. We sample $z$ and $\mathcal{M}$ using Markov Chain Monte Carlo (MCMC) with an ensemble sampler from the \textsc{emcee} package \citep{emcee}.  

\subsubsection{SFH prior}
\label{sec:priorscsfrd}

For the SFH prior, $p(\mathbf{x}_{\mathrm{sfr}}|z, \mathcal{M})$ we use the prescription used for the  \textsc{Prospector-}$\beta$ prior \citep{wang2023}, which sets the expected values for $\mathbf{x}_{\mathrm{sfr}} $ to the CSFRD from \citet{behroozi2019}. This prior also introduces an extra dependence on mass, to encode the expectation that high mass galaxies tend to form earlier in the history in the universe, and low mass galaxies later \citep{wang2023}.

However, we modify this approach to include uncertainties in the observational data used to infer the CSFRD in \citet{behroozi2019}. We train a GP model on the observational data compiled in \citet{behroozi2019}, such that we can draw many different realisations consistent with observations. This way, the expectation of $\mathbf{x}_{\mathrm{sfr}} $ is set to a particular realisation of CSFRD drawn from the GP model. We also subtract the training data by the `observed' fit shown in Figure 3 (Left) of \citet{behroozi2019}, to fix the mean of the GP.

\begin{figure}
    \centering
    \includegraphics[width=0.95\linewidth]{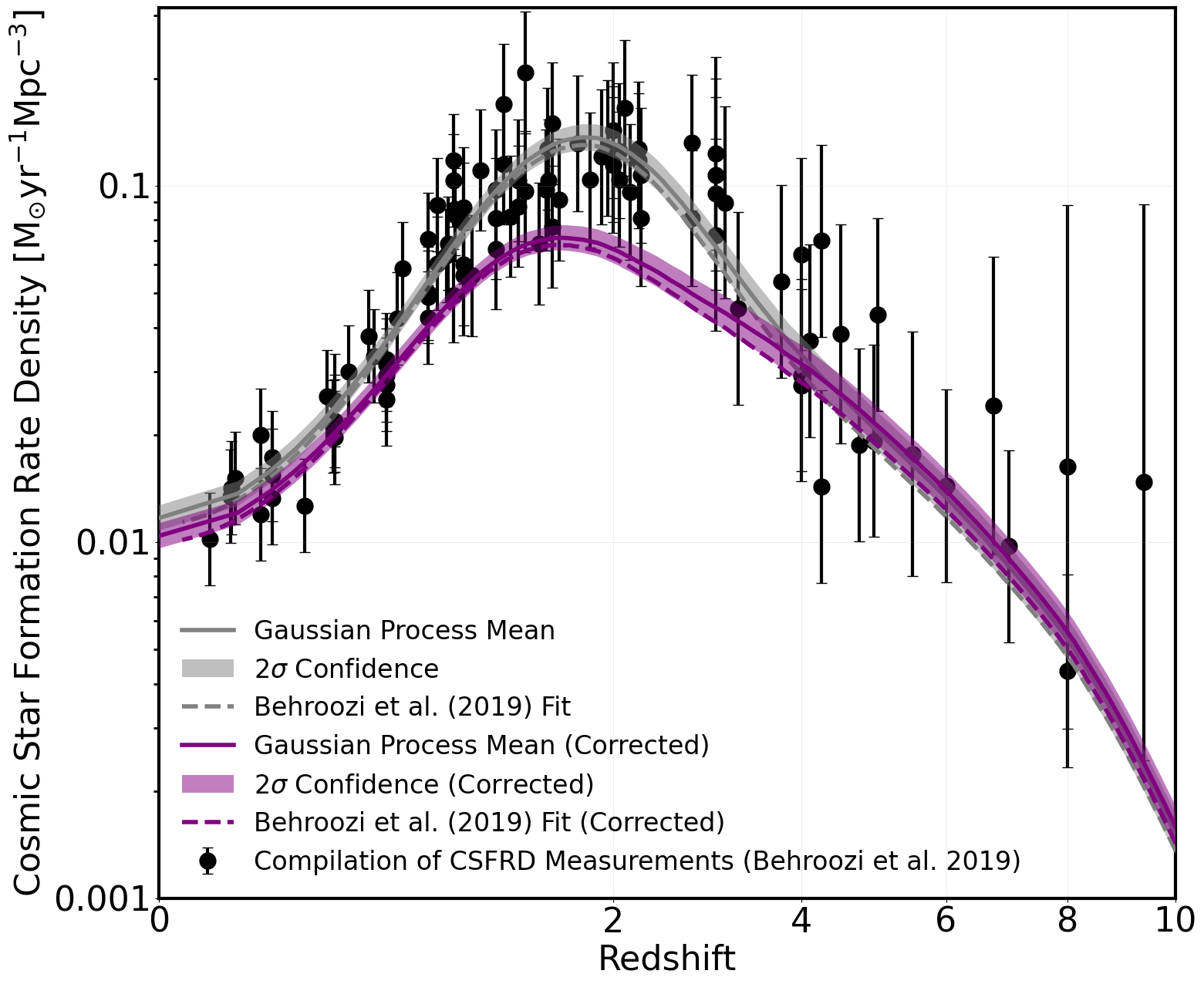}
    \caption{Distribution of CSFRD models sampled from the Gaussian process fit to observational measurements of the CSFRD compiled by \citet{behroozi2019}. The grey solid line show the Gaussian process mean to data shown by the black circles, while the purple solid line is the Gaussian process mean and after correcting for systematics. The shaded areas showing two standard deviations above and below the mean. The dashed lines show the inferred fits from \citet{behroozi2019}. The error bars have been made symmetric by taking the average of the upper and lower bound in \citet{behroozi2019}.}
    \label{fig:csfrd}
\end{figure}

Mirroring the approach in Section~\ref{sec:priorredmass}, \citet{behroozi2019} provides a training set of CSFRD measurements $\boldsymbol{\rho}_{\mathrm{obs}}$ at a set of observed redshifts $\mathbf{z}_{\mathrm{obs}}$, seen in Figure~\ref{fig:csfrd}. The GP provides a model for sampling the underlying noiseless evolution of the CSFRD $\boldsymbol{\rho}$, on a grid of redshifts $\mathbf{z}$ such that
\begin{equation}
    \boldsymbol{\rho} \sim p(\boldsymbol{\rho}|\boldsymbol{\rho}_{\mathrm{obs}}, \mathbf{z}, \mathbf{z}_{\mathrm{obs}}, \mathbf{C}_{\mathrm{obs}}^{\rho}),
    \label{eq:drawrho}
\end{equation}
where $\mathbf{C}_{\mathrm{obs}}^{\rho}$ characterises the observational noise on data $\boldsymbol{\rho}_{\mathrm{obs}}$, which is shown with the resulting GP fit in Figure~\ref{fig:csfrd}. We use an RBF kernel for $\mathbf{K}$ (equation \ref{eq:gpkernel}) as done in Section \ref{sec:priorredmass}. Additionally, we impose a prior such that $\sigma_{\ell} > 1.0$. We account for the systematic corrections in \citet{behroozi2019} by subtracting draws of the CSFRD in Equation~\ref{eq:drawrho} by the difference between the `observed' and `true' models in \citet{behroozi2019}. The distribution in Equation~\ref{eq:drawrho} with and without systematic correction is also shown in Figure~\ref{fig:csfrd}. We use the corrected CSFRD model for our $p(\mathbf{x}_{\mathrm{sfr}}|z, \mathcal{M})$ prior in the population model in Equation~\ref{eq:popmodelfac} following \citet{wang2023}.

\subsubsection{Dust prior}
\label{sec:priorsdust}

For our dust prior, $p(\boldsymbol{\psi}_{\mathrm{d}}|\mathbf{x}_{\mathrm{sfr}})$, we use samples from \textsc{pop-cosmos} \citep{popcosmos}. We draw 1,500,000 samples of $\boldsymbol{\psi}_{\mathrm{d}}$ and $\mathbf{x}_{\mathrm{sfr}}$ from the model, where we exploit the factorisation:
\begin{equation}
   p(\boldsymbol{\psi}_{\mathrm{d}}|\mathbf{x}_{\mathrm{sfr}}) = p(\delta|\tau_{2})p(\tau_{1}|\tau_{2})p(\tau_{2}|\mathbf{x}_{\mathrm{sfr}}).
\end{equation}
To sample $\boldsymbol\psi_{d}$ we first calculate the recent SFR averaged over the last 100~Myr, denoted by $\mathrm{SFR}_{100}$ and $\mathrm{SFR}'_{100}$ from our prior draws of $\mathbf{x}_{\mathrm{sfr}}$ (Section~\ref{sec:priorscsfrd}) and the \textsc{pop-cosmos} SFR ratios $\mathbf{x}'_{\mathrm{sfr}}$ respectively. To sample $\tau_{2}$, we linearly interpolate between the values diffuse dust parameters in the \textsc{pop-cosmos} sample $\tau'_{2}$ at $\mathrm{SFR}'_{100}$ to the SFR$_{100}$ drawn from our prior. This encodes the conditional distribution $p(\tau_{2}|\mathbf{x}_{\mathrm{sfr}})$ implied by \citet{popcosmos} for our draws of $\tau_{2}$. We then repeat the process to sample $\delta$ and $\tau_{1}$ using the linearly interpolated $\tau_{2}$, allowing us to draw from $p(\delta|\tau_{2})$ and $p(\tau_{1}|\tau_{2})$ implied by \citet{popcosmos}. We assume that these relations remain unchanged for $z>3$. This assumption is discussed in Section~\ref{sec:discussions}.

\subsubsection{Prior on metallicity, ionisation and AGN parameters}
\label{sec:priorselse}

The joint prior $P(U)P(Z)P(Z_{\mathrm{gas}})P(f_{\mathrm{agn}})P(\tau_{\mathrm{agn}})$ is assumed to be a product of Gaussians, with means and standard deviations left as free parameters. The means are sampled uniformly within prior bounds shown in Table~\ref{tab:spsparams}, with standard deviations sampled uniformly between 0.01 and the width of these prior bounds. This is summarised in Table~\ref{tab:popparams}. This uninformative and flexible prior is chosen as these parameters are found to have a limited effect on the simulated LBG redshift distributions compared to the SFH, redshift, stellar mass and the dust attenuation model parameters. While this means we sample potentially unphysical parts of parameter space, this choice will only make the forecast more conservative, as we will be marginalising over a larger parameter space than is realistic for these parameters. This is to avoid needing to extrapolate known relations at low redshift to $z>3$. For example,  \citet{alsing2023} take $Z_{\mathrm{gas}}$ to be conditioned on the SFH of the galaxy, which is observed to be broadly redshift invariant up to $z\sim3$ -- known as the fundamental metallicity relation (FMR) \citep{mannucci2010, cresci2019, curti2020}.  There is, however, evidence of evolution for $z>3$, including from recent observations using the James Webb Space Telescope (JWST), e.g., \cite{curti2024}. So we leave modelling an appropriate metallicity prior from $z=$0--7 based on observations to future work.

\subsection{Photometric noise model}
\label{sec:photonoise}

The SPS model detailed in Section \ref{sec:spsmodel} simulates noiseless photometry. To add noise we apply the LSST photometric noise model implemented in the package \textsc{Photerr}\footnote{https://github.com/jfcrenshaw/photerr} \citep{crenshaw2024} assuming 10 years of observation time. We also use this package to apply signal to noise (SNR) cuts by comparing the simulated noisy magnitudes to their respective LSST limiting magnitude for a given confidence level. These are listed in Section~\ref{sec:selec}.

\subsection{Selection}
\label{sec:selec}

Given a set of simulated noisy photometry, we then identify (candidate) LBGs using the dropout selection technique \citep{g1990, steidel96, giava2002}. For LSST we expect large numbers of galaxies that dropout in LSST $u$, $g$, and $r$ bands \citep{wilson19}. This means that for a given realisation of the population model, we apply three sets of selection cuts, one for dropouts in each band, giving three redshift distributions. 

Two types of selection cuts are used: SNR cuts and colour cuts. For the SNR cuts, we follow the procedure used for the GOLDRUSH LBG catalogue using Subaru-HSC \citep{goldrush4}. This requires that the $g$ and $r$ dropouts have SNR > 5 in the $i$ and $z$ bands respectively. Typically $u$ dropouts are detected in the $r$ band, however the $u$ band is much shallower, so instead of applying a SNR > 5 cut, we require $r > 25.7$. This is the 5$\sigma$ limiting magnitude of the $u$ band, where we assume LSSTY10 5$\sigma$ limiting magnitudes of $m_{\mathrm{lim}}=\{ 25.7,26.9,27.1,26.5,25.8\}$ for the $u, g, r, i$ and $z$ bands respectively, 
as given by \textsc{Photerr} \citep{crenshaw2024}. In addition, a further brightness cut is applied for all $u$, $g$, and $r$ dropouts, excluding sources brighter than a magnitude of 20 in $r$, $i$ and $z$ bands, respectively.

Following the SNR cuts above, we apply the LBG colour cuts using the $u$, $g$, $r$, $i$ and $z$ band magnitudes, where for the $u$ dropouts we use the colour cuts suggested by \citet{sawicki2019}:
\begin{align}
          (u-g) &> 0.88; \\
          (g-r) &< 1.2; \\
          (u-g) &>1.8(g-r)+0.68.
    \label{eq:ucut}
\end{align}
For the $g$ dropouts the colour cuts are \citep{goldrush4}:
\begin{align}
     (g-r) &> 1.0;\\  
     (r-i) &< 1.0; \\ 
     (g-r) &> 1.5(r-i) + 0.8.
 \label{eq:gcut}
\end{align}
Finally for the $r$ dropouts \citep{goldrush4}:
\begin{align}
 (r-i) &> 1.2;\\ 
 (i-z) &< 0.7; \\ 
 (r-i) &> 1.5(i-z) + 1.0.
 \label{eq:rcut}
\end{align}

These colour cuts help bin $u$, $g$, and $r$ dropouts at $z\sim3, 4$ and 5 respectively, while minimising contaminants such as low $z$ interloper galaxies and stars \citep{giava2002, shap11, cars09, wilson19, goldrush4}. The strategy for LBG selection for LSST is not yet finalised, so while the selection cuts are inspired by methodology in the literature \citep{cars09, sawicki2019, goldrush4}, the actual selection cuts used will likely be different. There is even the possibility of replacing traditional colour cuts with more sophisticated methods, which utilise more of the available colour-colour space \citep{payerne25, crenshaw2025}. 

\subsection{SPS emulation}
\label{sec:emulation}
Simulating large numbers of galaxy SEDs is computationally expensive. For this forecast not only do we need simulate a large enough sample of galaxies to construct the redshift distributions, but also repeat this process many times to sample multiple realisations of the galaxy population model. 

Therefore we make the use of SPS emulators to speed up the simulations of galaxy SEDs. We use \textsc{Speculator} \citep{speculator} to build and train five neural networks (one for each of LSST bands) to emulate our SPS photometry. These are trained on a  set of $1.2\times10^8$ galaxies with SPS parameters sampled uniformly between the prior bounds defined in Table \ref{tab:spsparams}. This is a much larger training set than used in \citet{speculator}, as extending the redshift range to $z=7$ makes training considerably more difficult. To account for this, we use a larger network, with hidden layers containing the following number of neurons per layer: $[64, 128, 256, 256, 256, 256, 128, 64, 32, 16, 8, 4, 2]$. This is found to produce fluxes where the 5th and 95th percentiles are consistent with FSPS to within $5 \%$, across the training data. An error floor on fluxes of $\sim 5 \%$ is typically used to account for uncertainties in photometric calibration \citep{speculator}, so this is sufficient for our requirements.

\begin{figure*}
    \centering
    \includegraphics[width=0.95\textwidth]{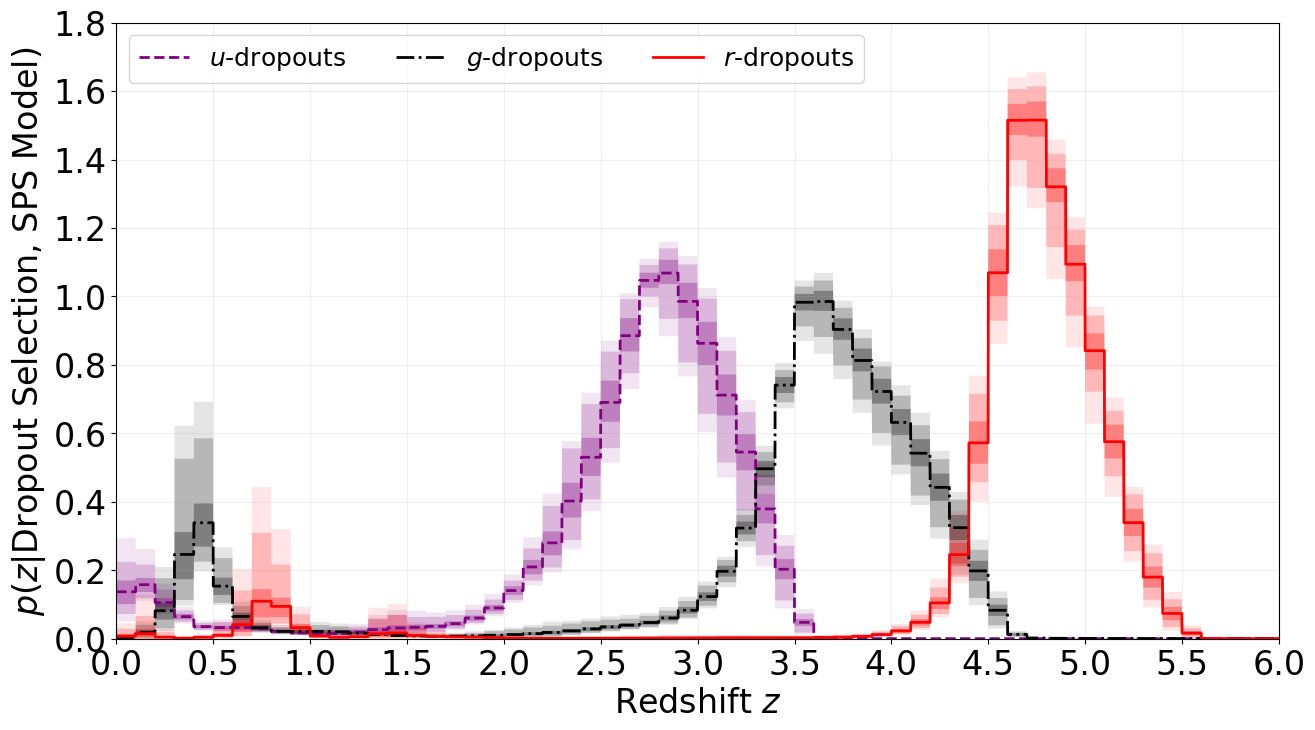}
    \caption{Prior on $u$ (purple), $g$ (black) and $r$ (red) normalised LBG dropout redshift distributions. The solid lines indicate the mean, with the shaded regions (dark to light) are the 16-84th, 2.5-97.5th and 0.3-99.7th percentiles respectively, showing the variation in the redshift distribution caused by samples of galaxies draw from different realisations of the galaxy population model.}
    \label{fig:simnz}
\end{figure*}

\begin{figure}
    \centering
    \includegraphics[width=0.95\linewidth]{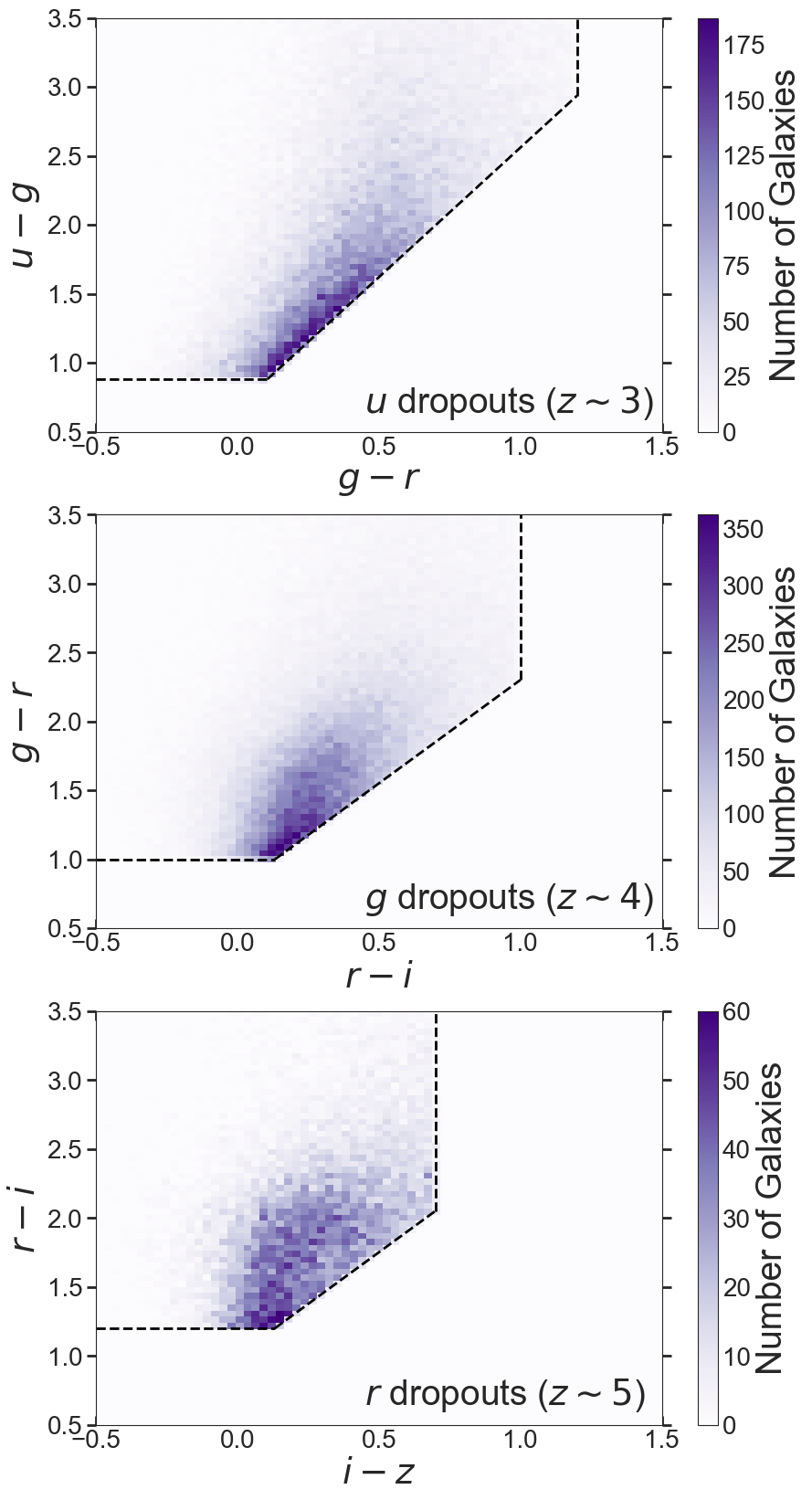}
    \caption{Colours of $u$, $g$ and $r$ dropout LBGs (top to bottom) selected by the cuts defined in Section \ref{sec:selec}, from an initial sample of 4,000,000 galaxies drawn from the mean of our population model. The dashed black lines show the LBG colour cuts described in the text.}
    \label{fig:coloucolour}
\end{figure}

\subsection{Redshift distributions}
\label{sec:nzs} 

We can use the forward model to compute redshift distributions by sampling different realisations of $\boldsymbol{\varphi}$ for the population model as described in Section~\ref{sec:popmodel}. This will allow us to quantify how uncertainties in the population model propagates to the LBG redshift distributions. This will also form the basis of the redshift distribution prior in our forecast in Section ~\ref{sec:comsoforecast}. 

We generate $n=1024$ realisations of $\boldsymbol{\varphi}$, where for each we simulate $N_{\mathrm{sim}} = 4 \times 10^{6}$ galaxies, resulting in $N_{\mathrm{sim}}$ sets of SPS parameters $\boldsymbol{\psi}$ for each population model realisation. These $\boldsymbol{\psi}$ are passed to the SPS model detailed in Section~\ref{sec:spsmodel} to produce noiseless photometry using the neural network emulator in Section~\ref{sec:emulation}. After applying the photometric noise model and selection cuts detailed in Section~\ref{sec:photonoise} and  \ref{sec:selec}, respectively, we generate a total of $n$ sets of three LBG redshift distributions, one for each dropout subpopulation $u$, $g$ and $r$. This is shown in Figure~\ref{fig:simnz}, with the colours for the population model mean shown in Figure~\ref{fig:coloucolour}. The redshift distributions are multimodal, showing the LBG dropout population at $z \gtrsim 2$, and the low redshift contaminants. These distributions are qualitatively consistent with previous observations using data from HSC, CFHT U-band Survey (CLAUDS) and DESI \citep{goldrush4, ruhlmann24, mons24}. The interloper distributions peak at redshifts consistent with Balmer-break/4000\AA \  contaminants, aside from a small population visible at $z \sim 1.4$ for the $r$ dropouts, which to our knowledge there is no evidence for in the literature. As the mean is very close to zero here, this may be an artifact from sampling remote regions of our population model parameter space that may be unphysical. 

\begin{figure}
    \centering
    \includegraphics[width=0.95\linewidth]{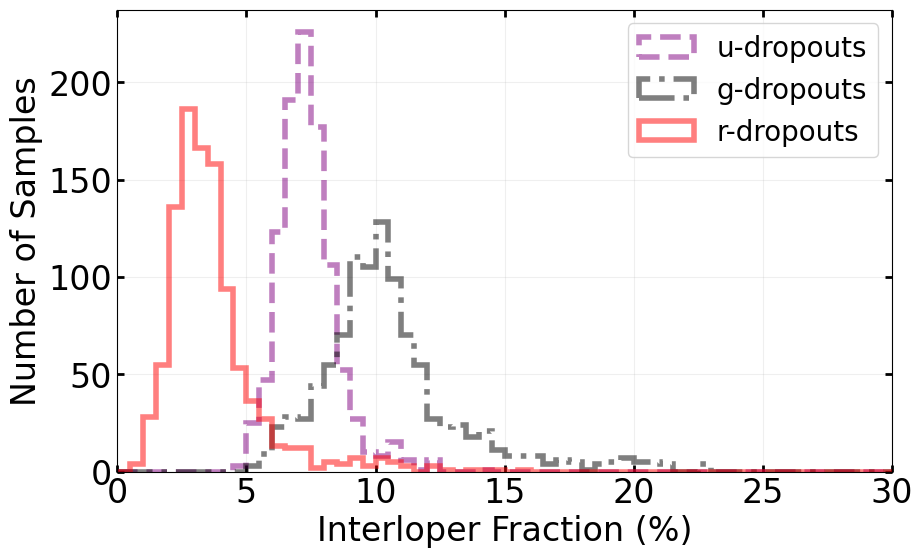}
    \caption{The distribution of interloper fractions (Defined in Section~\ref{sec:nzs}) for each dropout population, from 1024 realisations of the population model.}
    \label{fig:intfrac}
\end{figure}

The distribution of interloper contamination fractions $f_{\mathrm{int}}$, defined as the fraction of galaxies selected with $z < 1.5$, is shown in Figure~\ref{fig:intfrac}. This shows mean interloper fractions of $(7 \pm 1)\%$, $(11 \pm 3)\%$, and $(4 \pm 2) \%$ for the $u$, $g$ and $r$ dropouts respectively. We find a higher interloper fraction for the $g$ dropouts compared to $(2 \pm 3) \%$ in \citet{goldrush4}, which is estimated for the HSC deep and ultra deep layers. These have comparable limiting magnitudes to our analysis. However, our results for the $r$ dropouts are consistent with the $\sim$1\%--9\%  fraction estimated in \citet{goldrush4}. The uncertainties in our predicted $f_{\mathrm{int}}$ are due to different realisations of the population model introduced by sampling parameters $\boldsymbol{\varphi}$, where the errors quoted show one standard deviation from the mean. The interloper contamination fraction is sensitive to the selection cuts used, and there may be room to optimise these cuts for LSST to reduce these if needed \citep{ruhlmann24, payerne25}.

Given the mass functions used in our population prior (Section~\ref{sec:priorredmass}, we can also estimate the number densities of a particular LBG dropout subpopulation $d$ (i.e. $u$, $g$ or $r$ dropouts) 
observed by LSST as
\begin{equation}
    \bar{n}_{d} = \frac{1}{\Omega}\int^{\infty}_{0} \int_{\mathcal{M}_{\mathrm{lim}}(z)}^{\infty} p(S_{d}|\mathcal{M}, z)\Phi(\mathcal{M}, z)\frac{dV_{\mathrm{co}}}{dz}d\mathcal{M}dz,
    \label{eq:lbgnumden}
\end{equation}
where $p(S_{d}|\mathcal{M}, z)$ is the probability of detecting a galaxy with given $\mathcal{M}$ and $z$ with selection cuts $S_{d}$ (Section \ref{sec:selec}) and $\Omega$ is the total sky area in steradians. 
The detection probability can be estimated from the simulations as
\begin{equation}
    p(S_{d}|\mathcal{M}, z) \approx\frac{N_{\mathrm{sim},d}(\mathcal{M}, z)}{N_{\mathrm{sim}}(\mathcal{M}, z)},
\end{equation}
where $N_{\mathrm{sim},d}(\mathcal{M}, z)$ is the total number of dropouts passing the selection cuts. The lower prior bound for the log mass is shown in Table \ref{tab:spsparams} as $\mathcal{M} = 7$, so we set $\mathcal{M_{\mathrm{lim}}} = 7$. To assist with efficient sampling of SPS parameters, we further impose $\mathcal{M_{\mathrm{lim}}} = 8$ for $z > 1$, to avoid simulating galaxies are too faint to be detected. This has a negligible effect on the calculated number densities. For this comparison we have assumed the limiting magnitudes in the respective detection bands for each dropout as 25.7, 26.5 and 25.8 as described in Section \ref{sec:selec}. 

The implied number densities of LBGs (including low redshift interlopers) are $n_{\mathrm{u}} = (8000 \pm 1000) \ \mathrm{deg}^{-2}$, $n_{\mathrm{g}} = (14000 \pm 2000)\ \mathrm{deg}^{-2}$ and $n_{\mathrm{r}} = (1100 \pm 400) \ \mathrm{deg}^{-2}$, for $u$, $g$ and $r$ dropouts, respectively. These uncertainties arise from the flexibility introduced into the population model via parameters $\boldsymbol{\varphi}$, where the errors quoted are one standard deviation from the mean. The predicted number densities are broadly consistent with the number densities in Figure~6 of \citet{wilson19}, which predicts $\sim 10000\ \mathrm{deg}^{-2}$ $u$ dropouts, $\sim 10000\ \mathrm{deg}^{-2}$ $g$ dropouts and $\sim 1000\ \mathrm{deg}^{-2}$ $r$ dropouts. Our approach allows us to improve upon the estimates in \citet{wilson19} as we can begin to quantify an uncertainty on the expected number densities, from the expected uncertainty in the galaxy population prior.

\section{Cosmological Forecast}
\label{sec:comsoforecast}

In this section we will show how we can use the simulated SPS redshift distributions in Section \ref{sec:forwardmodel}, and their associated uncertainties resulting from the flexibility introduced in the population model, as a redshift distribution prior for a forecast on cosmological parameters. This forecast will give an insight into the ability of the LSSTY10 LBG photometry to constrain cosmological parameters, without any spectroscopic information. 

In Section~\ref{sec:marginalisation} we will show how we marginalise over the redshift distribution uncertainties, followed by a discussion of the parameterisation of the redshift distributions in Section~\ref{sec:pca}. Finally we show a Fisher forecast on cosmological parameters for an LSSTY10 style survey in Sections~\ref{sec:fisher} $\&$ \ref{sec:results}. We will forecast constraints on $\sigma_{8}$, the baryonic and dark matter energy density parameters $\Omega_{c}$ and $\Omega_{b}$ respectively, $h$, and the spectral index $n_{\mathrm{s}}$. These will be evaluated at a `Planck 2015' cosmology \citep{planck15}, which assumes $\{ \sigma_{8}=0.816, \Omega_{\mathrm{c}}=0.259, \Omega_{\mathrm{b}}=0.0486, h=0.677, n_{\mathrm{s}}=0.967\}$. The dark energy equation of state parameter is left fixed at $w=-1$ as we are probing the universe during matter domination. We will also provide constraints on nuisance parameters for the LBG galaxy bias.

\subsection{Marginalisation over the redshift distributions}
\label{sec:marginalisation} 

In this section we will briefly detail the redshift distribution marginalisation calculation presented in \citet{hadz2020}.

The simulated redshift distributions shown in Figure \ref{fig:simnz} include an uncertainty introduced by variations in the population model used. Different galaxy population models result in different LBG redshift distributions, which have an affect on inferred cosmological parameters. Therefore, we can use the redshift distributions presented in Section \ref{sec:forwardmodel} as a prior, which can be marginalised over to propagate the population model uncertainties to cosmological parameters.

The quantity of interest is the posterior distribution, $p(\boldsymbol\theta | \mathbf{d})$,  of the model parameters, $\mathbf{\boldsymbol\theta}$, conditioned on the data. For this analysis $\boldsymbol\theta  = (\mathbf{q}, \mathbf{N})$, where $\mathbf{q}$ are the cosmological and nuisance parameters and $\mathbf{N}$ are parameters describing the galaxies' redshift distribution. Applying Bayes theorem yields
\begin{equation}
    p(\mathbf{q}, \mathbf{N} | \mathbf{d}) \propto p(\mathbf{d} | \mathbf{q},  \mathbf{N})p(\mathbf{q})p(\mathbf{N}),
    \label{eq:bayes}
\end{equation}
where $p(\mathbf{d} | \mathbf{q},  \mathbf{N})$ is the likelihood, and $p(\mathbf{q})$ and $p(\mathbf{N})$ are the priors. The likelihood is assumed to be a multi-variate normal, given by 
\begin{align}
    p(\mathbf{d} | \mathbf{q},  \mathbf{N}) &\propto \exp{\bigg(-\frac{1}{2}(\mathbf{d} - \mathbf{m(q, N)})^{T}\mathrm{\mathbf{C}_{c}^{-1}}(\mathbf{d} - \mathbf{m(q, N)})\bigg)}, \\
    &=\exp{\big(-\frac{1}{2} \chi_{c}^{2}\big)},
\end{align}
which implicitly defines $\chi_{c}^{2} \equiv (\mathbf{d} - \mathbf{m(q, N)})^{T}\mathrm{\mathbf{C}_{c}^{-1}}(\mathbf{d} - \mathbf{m(q, N)})$, $\mathbf{m} = \mathbf{m(q, N)}$ is the prediction for $\mathbf{d}$, and $\mathbf{C}_{c}$ is the data covariance matrix. The model vector gives the prediction for the angular power spectrum in Equation~\ref{eq:angpower}, for a given set of $u$ , $g$, and $r$ dropout redshift distributions and the CMB lensing kernel given in Equation~\ref{eq:cmbkernel}.

The prior on the redshift distribution parameters is also assumed to be Gaussian, such that
\begin{align}
    p(\mathbf{N}) &\propto \exp{\bigg(-\frac{1}{2}(\mathbf{N} - \mathbf{\Bar{N}})^{T}\mathrm{\mathbf{P}^{-1}}(\mathbf{N} - \mathbf{\Bar{N}})\bigg)}, \\
    &=\exp{\big(-\frac{1}{2} \chi_{\mathrm{p}}^{2}\big)}
    \label{eq:redprior}
\end{align}
given $\chi_{\mathrm{p}}^{2} \equiv (\mathbf{N} - \mathbf{\Bar{N}})^{T}\mathrm{\mathbf{P}^{-1}}(\mathbf{N} - \mathbf{\Bar{N}})$, where $\mathbf{P}$ is the prior covariance and $\mathbf{\Bar{N}}$ is the mean of $\mathbf{N}$. Marginalising over $\mathbf{N}$ in Equation~\ref{eq:bayes} we get
\begin{align}
    p(\mathbf{q} | \mathbf{d}) &\propto p(\mathbf{q}) \int p(\mathbf{d} | \mathbf{q},  \mathbf{N})p(\mathbf{N}) d\mathbf{N} \\
     &= p(\mathbf{q}) \int \exp{(-\frac{1}{2} [ \chi_{\mathrm{c}}^{2} + \chi_{\mathrm{p}}^{2}  ])} d\mathbf{N},
    \label{eq:marg}
\end{align}
which needs to be evaluated to obtain constraints on $\mathbf{q}$. However, the integral in Equation~\ref{eq:marg} becomes computationally intensive to calculate with a large number of redshift distribution parameters for $\mathbf{N}$. This is particularly true when parameterising the distributions as histograms as in Figure \ref{fig:simnz} or for a many component PCA decomposition, which we will introduce in Section~\ref{sec:pca}.

Instead, we perform the marginalisation analytically, using the method described in \citet{hadz2020}. Briefly, this method works by expanding the model vector to linear order, yielding
\begin{equation}
    \mathbf{m(q, N)} \simeq \mathbf{m(q, \Bar{N})} + \mathbf{T}(\mathbf{N}-\mathbf{\Bar{N}}),
    \label{eq:margapprox}
\end{equation}
where 
\begin{equation}
    \left.\mathbf{T} \equiv \frac{d\mathbf{m}}{d\mathbf{N}}\right|_{\mathbf{q},\mathbf{ \Bar{N}}}
\end{equation}
 encodes the response of the predicted power spectra to changes in the redshift distribution parametrised by $\mathbf{N}$. By applying Equation~\ref{eq:margapprox} to the integral in Equation~\ref{eq:marg}, \citet{hadz2020} obtains a approximation to the marginalised likelihood as
\begin{multline}
    \label{eq:marganalytical}
    p(\mathbf{d} | \mathbf{q}) \propto \\
    \sqrt{\mathrm{det}(\mathbf{T}^{T}\mathbf{C}_{\mathrm{c}}^{-1}\mathbf{T}+\mathbf{P}^{-1})} \exp{[\frac{1}{2}(\mathbf{d} - \mathbf{m})^{T}\mathbf{C}_{\mathrm{m}}^{-1}(\mathbf{d} - \mathbf{m})]},
\end{multline}
where 
\begin{equation}
    \mathbf{C}_{\mathrm{m}} = \mathbf{C}_{\mathrm{c}} + \mathbf{T}\mathbf{P}\mathbf{T}^{T}.
    \label{eq:cm}
\end{equation}
If $\mathbf{T}$ does not depend significantly on $\mathbf{q}$, it can be kept it fixed in Equation~\ref{eq:marganalytical} and  $p(\mathbf{d} | \mathbf{q})$ can be approximated as a multi-variate normal with an inflated covariance $\mathbf{C}_{\mathrm{m}}$ due to the marginalisation over $\mathbf{N}$. 

\subsection{Parameterisation of the redshift distributions}
\label{sec:pca} 

The marginalisation scheme in Section~\ref{sec:marginalisation}, requires that the redshift distribution be specified using parameters that are approximately Gaussian distributed (Equation~\ref{eq:redprior}). One choice of parameterisation would be histogram bin heights; however, as seen in Figure~\ref{fig:simnz}, in some places these are heavily skewed if close to zero (i.e., the low redshift interlopers).

Instead, we chose a parameterisation for $\mathbf{N}$ by performing principle component analysis (PCA) on the sample of simulated redshift distributions. This choice is motivated by the fact that PCA assumes Gaussian latent space variables that we can use for our parameterisation. The forward model detailed in Section \ref{sec:forwardmodel} produces simulated redshift distributions as histograms with equal bin widths. These are parametrised by a set of bin heights such that $\mathbf{N} = (h_{1}, h_{2} , ..., h_{B} )^{T}$, where $h_{b}$ is the height of bin $b$, and $B$ is the total number of bins. For the simulated redshift distributions seen in Figure~\ref{fig:simnz} we have a total of $B=70$ bins of width 0.1 between 0 and 7. To ensure that $N(z) \geq 0 \ \forall \ z$ we perform the PCA decomposition on $\sqrt{h_{b}}$ for the set of $n$ simulated redshift distributions. We fit an $n_{\mathrm{p}}$ component PCA such that we reduce $\mathbf{N}$ to $\mathbf{N}_{\mathrm{pca}} = (a_{1}, a_{2},\  ...\  a_{n_{\mathrm{p}}})^{T}$, which are the new redshift distribution parameters. We choose $n_{\mathrm{p}}$ = 50, as this gives the total explained variance ratio of at least 0.95 for each of the dropouts. 
The parameters $\mathbf{N}_{\mathrm{pca}}$ are the coefficients to the set of $n_{\mathrm{p}}$ PCA eigenvectors $\nu_{\mathrm{p}} = \{ \mathbf{v}_{1}, \mathbf{v}_{2} \ ...\  \mathbf{v}_{n_{\mathrm{p}}} \}$ recovered from the fit, where each are of length $B$.  Using this reduced parameterisation, we can recover the implied bin heights as
\begin{equation}
h_{b} = \bigg( \bar{h}^{\mathrm{pca}}_{b} + \sum_{l=1}^{n_{\mathrm{p}}}a_{l}v_{lb} \bigg)^{2},
    \label{eq:pcaeqn}
\end{equation}
where $v_{lb}$, are the components of the $B \times n_{\mathrm{p}}$ matrix $\mathbf{V}$, where the columns of the matrix are given by the elements of $\nu_{\mathrm{p}}$. The quantity  $\bar{h}^{\mathrm{pca}}_{b}$ is square root of the mean height of bin $b$ over the training set. The expression on the right-hand side of Equation~\ref{eq:pcaeqn} is squared, this is because we have fitted the 50-component PCA to $\sqrt{h_{b}}$ instead of $h_{b}$.
We can sample $\mathbf{N}_{\mathrm{pca}}$, and therefore the redshift distributions, using this PCA approximation by drawing
\begin{equation}
    \mathbf{N}_{\mathrm{pca}} \sim \mathcal{N}(\mathbf{0}, \mathbf{C}^{d}_{\mathrm{pca}}).
    \label{eq:drawpca}
\end{equation}
The covariance matrix $\mathbf{C}^{d}_{\mathrm{pca}}$ is a $50 \times 50$ diagonal matrix containing the eigenvalues from the PCA fit, for a given dropout $d$. This PCA parameterisation approximates the non-Gaussian distribution of the correlated bin heights as a multivariate Gaussian via Equation~\ref{eq:drawpca}. Therefore we can exploit the analytical marginalisation procedure detailed in Section~\ref{sec:marginalisation} as our new redshift distribution parameterisation given by parameters $\mathbf{N}_{\mathrm{pca}}$ is Gaussian distributed.

\subsection{Fisher formalism and assumptions}
\label{sec:fisher}
To obtain predicted constraints on cosmological parameters using the redshift distribution model in Section \ref{sec:forwardmodel}, we calculate the Fisher matrix for our likelihood in equation \ref{eq:marganalytical}. By keeping $\mathbf{T}$ fixed, the likelihood is gaussian with a fixed covariance, so we may calculate the marginalised Fisher matrix $\mathbf{F}_{\mathrm{marg}}$ as \citep{tegmark1997}
\begin{equation}
    (\mathbf{F}_{\mathrm{marg}})_{q_{i}q_{j}} = \sum_{\alpha\beta} \frac{\partial m_{\alpha}}{\partial q_{i}}(\mathbf{C}^{-1}_{m})_{\alpha\beta}\frac{\partial m_{\beta}}{\partial q_{j}},
    \label{eq:fisher}
\end{equation}
where $m_{\alpha}$ and $q_{i}$ are components of $\mathbf{m}$ and $\mathbf{q}$, respectively. Constraints on $q_{i}$ can be estimated by calculating $\mathbf{F}_{\mathrm{marg}}^{-1}$ \citep{tegmark1997}.
Treating $\mathbf{T}$ as being independent of $\mathbf{q}$ has been shown to have little impact on inferred posteriors over numerical marginalisation \citep{hadz2020, ruiz2023}. For this reason, combined with the approximate nature of Fisher forecasts, it is reasonable to assume that keeping $\mathbf{T}$ fixed will have a negligible impact on the constraints obtained.

The model vector $\mathbf{m}(\mathbf{q}, \mathbf{N})$ contains the prediction of the angular power spectrum for the $u$, $g$ and $r$ dropouts from equation \ref{eq:angpower}, including cross-spectra, and cross-correlations with the CMB lensing following \citet{schaan2020, schmittfull2018}. LBGs are seen mainly at higher redshift (z $\gtrsim$ 2), a linear bias model should be sufficient up to $\ell \sim 1000$ \citep{wilson19}. Hence, we apply scale cuts such that each power spectrum is evaluated $200 \leq \ell \leq 1000$, with the reasoning behind the large-scale cut discussed in Section~\ref{sec:intbias}. Therefore, $\mathbf{m}$ is a concatenated vector of 10 angular power spectra, with a total of $10 \times 800 = 8000$ components, meaning $\mathbf{C}_{c}$ has $8000 \times 8000$ entries. The covariance is assumed to to have contributions from cosmic variance and observed sky fraction $f_{\mathrm{sky}}$ only following \citet{schmittfull2018, schaan2020}, where we include a Poisson noise contribution to the auto-power spectra. For this work we do not model CMB lensing reconstruction noise, as noise on $C_{\ell}^{\kappa \kappa}$ should be mostly cosmic variance limited within $200 \leq \ell \leq  1000$ \citep{schaan2020}.

The prior covariance $\mathbf{P}$ in matrix block notation is
\begin{equation}
\mathbf{P} = 
\begin{bmatrix}
    \mathbf{C}^{\mathrm{u}}_{\mathrm{pca}} & \mathbf{0} & \mathbf{0} \\ 
    \mathbf{0} & \mathbf{C}^{\mathrm{g}}_{\mathrm{pca}} & \mathbf{0} \\ 
    \mathbf{0} & \mathbf{0} & \mathbf{C}^{\mathrm{r}}_{\mathrm{pca}},
    \label{eq:priorcov}
\end{bmatrix}
\end{equation}
where $\mathbf{0}$ is a $50\times50$ zero matrix. Each redshift distribution is described by 50 PCA parameters, so with three dropouts, $\mathbf{P}$ is a $150 \times 150$ matrix. We also assume a linear bias model, parametrised by $b_{\mathrm{u}}$, $b_{\mathrm{g}}$ and $b_{\mathrm{r}}$ for the $u$, $g$ and $r$ dropouts respectively as  introduced in  Section~\ref{sec:background}.

\subsection{Forecast}
\label{sec:results}

We perform a Fisher forecast for an LSSTY10 style data set with the formalism laid out in Section~\ref{sec:fisher}, given $f_{\mathrm{sky}} = 0.35$. To evaluate the derivatives in Equation~\ref{eq:fisher} easily, we use \textsc{JAX-COSMO} \citep{jaxpaper} for calculating angular power spectra, within the Limber approximation \citep{limber53}. This allows us to leverage \textsc{JAX} auto-differentiation for calculating derivatives quickly, without the need for finite-differences. 
The Fisher matrix is calculated for the set of parameters $\mathbf{q} = (\sigma_{8}, \Omega_{\mathrm{c}}, \Omega_{\mathrm{b}}, h, n_{\mathrm{s}}, b_{\mathrm{u}}, b_{\mathrm{g}}, b_{\mathrm{u}})$, where $b_{\mathrm{u}} = 3, b_{\mathrm{g}} = 4 $ and $ b_{\mathrm{r}} = 5$. The bias parameters are evaluated at these values to approximately mirror the redshift dependent bias of LBGs seen in a collection of measurements summarised in Figure~7 of \citet{wilson19}. We assume mean LBG number densities predicted by the model in Section~\ref{sec:nzs} for the treatment of Poisson noise. Alongside $\mathbf{F}_{\mathrm{marg}}$, we also calculate another Fisher matrix $\mathbf{F}_{0}$. We evaluate Equation \ref{eq:fisher} instead for a fixed redshift distribution, at the mean $\mathbf{N} = \mathbf{N}_{\mathrm{pca}} = \mathbf{0}$, such that $\mathbf{C}_{\mathrm{m}}=\mathbf{C}_{\mathrm{c}}$.

\begin{figure*}
    \centering
    \includegraphics[width=1.0\textwidth]{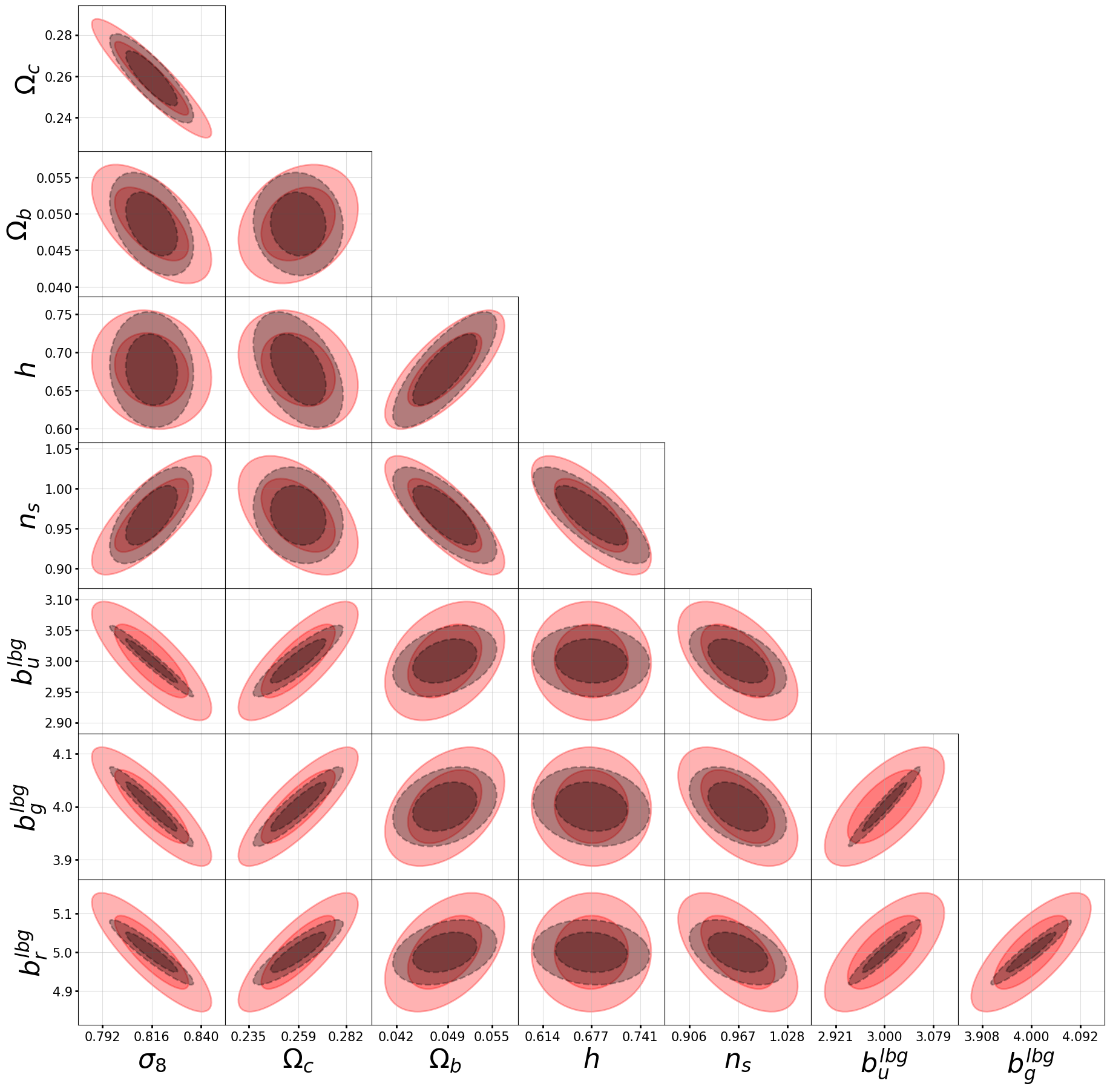}
    \caption{Forecast on cosmological and nuisance bias parameters, marginalising over modelled redshift distribution uncertainties (red-solid), and assuming a fixed redshift distribution (black-dashed). Contours show 68\% and 95 \% probability enclosed.}
    \label{fig:trig_plot}
\end{figure*}

The resulting forecast constraints are compared in Figure~\ref{fig:trig_plot}.
It is clear that the extra redshift distribution space explored by the SPS model significantly reduces the constraining power, particularly for $\sigma_{8}$ and the bias parameters. The presence of the low-redshift interlopers in the prior (Figure~\ref{fig:simnz}) introduces a mixing of probed scales, which manifests itself as uncertainty in parameters such as $\sigma_{8}$, which is expected to vary significantly between $z\sim0.5$ and $z>3$. These interlopers, if not accounted for, result in biased inference of the cosmological parameters, as discussed further in \citet{wilson19}. 

\begin{figure}
    \centering
    \includegraphics[width=0.95\linewidth]{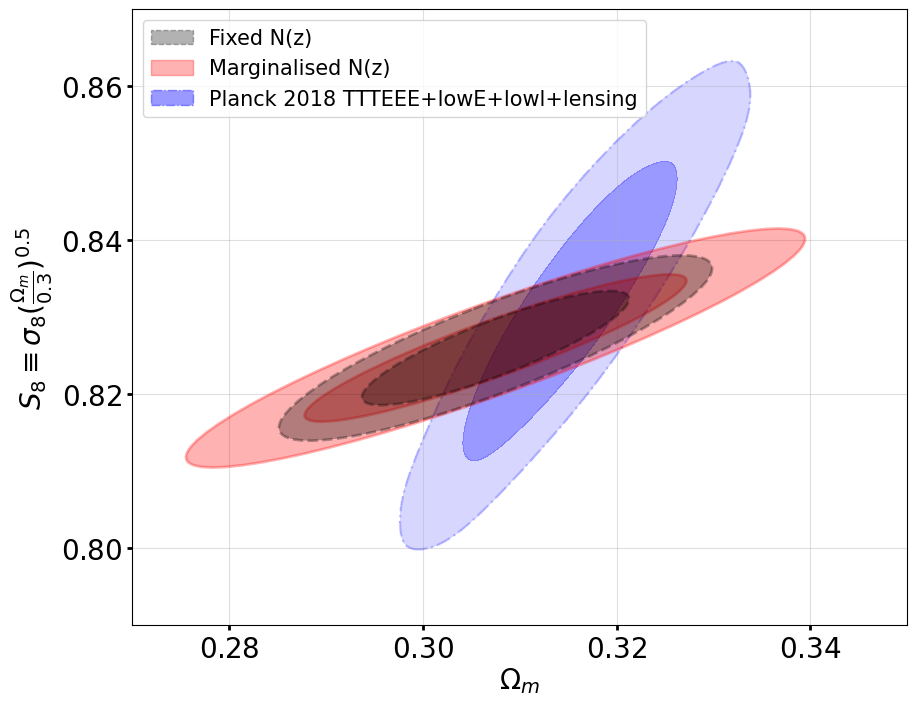}
    \caption{Forecast of $S_{8}$ and $\Omega_{m}$ parameters for a fixed, known redshift distribution (black-dashed), marginalised over redshift distribution uncertainties (red-solid). Constraints from Planck 2018 data \citep{planck18} (blue-dashed-dotted) are also shown. Contours show 68\% and 95 \% probability enclosed. The contours from our forecast are shifted slightly from the Planck results purely as a consequence of choosing to evaluate the Fisher matrix at a Planck 2015 cosmology, which is slightly different to the Planck 2018 results. Fisher forecasts only provide information on the precision of parameter constraints.}
    \label{fig:s8}
\end{figure}

Regardless, the marginalised constraints shown in Figure~\ref{fig:trig_plot} are promising, giving  precision on $\sigma_{8}$ comparable to Planck 2018 data \citep{planck18}. A comparison is shown in Figure~\ref{fig:s8}. These constraints are much more informative than can be achieved with previous surveys such as in \citet{miya22}. Our results show that accounting for redshift distribution uncertainties resulting from uncertainty in the galaxy population can yield similarly tight constraints on $\sigma_{8}$ as predicted by previous forecasts \citep{schmittfull2018, wilson19}. However our results are dependent a fixed galaxy population dust model, where we will investigate the impact of our choice in the following section.

\section{Discussion}
\label{sec:discussions}

\subsection{Uncertainty in the galaxy population dust model}

The choice to fix the dust model using results from \citet{popcosmos} is potentially significant. We found that simulated LBG redshift distributions are very sensitive to the parameters $\boldsymbol\psi_{\mathrm{d}}$, so a flexible parameterisation of the population dust prior $p(\boldsymbol{\psi}_{\mathrm{d}}|\mathbf{x}_{\mathrm{sfr}})$ was not feasible as with the other parameters. Also, available dust models are derived from observations of galaxies mostly outside our redshift range of interest, and show differing quantities of ISM dust.

The work in \citet{popcosmos} is derived mostly from data at redshifts $\lesssim 3$ (being based on a subset of COSMOS with $r<25$~mag), so the distribution of $\tau_{1}, \tau_{2}$ and $\delta$ may differ for higher-redshift galaxies than are probed by this selection. An updated \textsc{pop-cosmos} model from Thorp et al.\ (in prep.) is based on a deeper subsample of COSMOS (with \textit{Spitzer} IRAC Ch.\ 1 $<26$~mag; following \citealp{weaver2023}), which has non-negligible number counts at $z\lesssim5$, and probes a redder population of galaxies. We also show results from this updated \textsc{pop-cosmos} model in the comparisons that follow.

\begin{figure}
    \centering
    \includegraphics[width=0.95\linewidth]{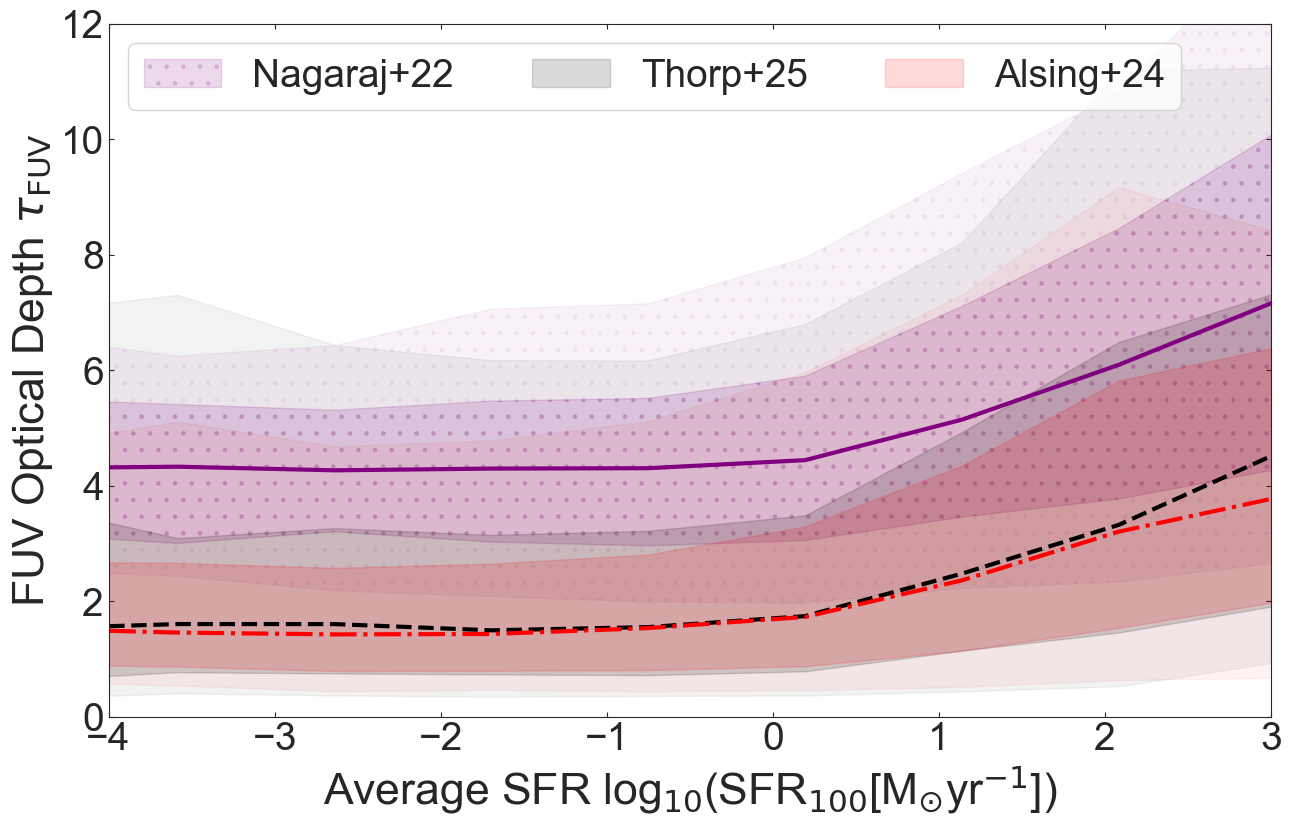}
    \caption{Comparison of the FUV dust optical depth as predicted by \citet{popcosmos} (black, dashed-dotted curve),  \citet{nagaraj2022} (purple, solid curve) and Thorp et al. (in prep.) (black, dashed curve). Samples drawn from these models are binned in SFR where the curves show the median, and the shaded areas show the 68\% and 95\% confidence intervals for each bin respectively.}
    \label{fig:dust_models}
\end{figure}

We find that another dust model, provided by \citet{nagaraj2022} based on data from the {\em Hubble Space Telescope} ({\em HST}) between $0.5 < z < 3.0 $ predicts galaxies with more dust on average than \citet{popcosmos}. Figure~\ref{fig:dust_models}  shows the median optical depth $\tau_{\mathrm{FUV}}$ in the rest frame far ultraviolet (evaluated at 1500~\AA), to be on average more than double that of \citet{popcosmos} as a function of $\mathrm{SFR}_{100}$ (the average of the SFR in the last 100~Myr). The \citet{popcosmos} and Thorp et al.\ (in prep.) results are similar to one another in their median attenuation vs.\ SFR$_{100}$ relations, with Thorp et al.\ (in prep.) predicting a slightly heavier tail towards high attenuation. It should be noted that here we estimate $\tau_{\mathrm{FUV}}$ from the samples by assuming that the majority of light emitted from galaxies is attenuated by both components of the dust model, due to the nature of LBG SEDs being dominated by light emitted by young stars in the rest frame ultraviolet.

\begin{figure}
    \centering
\includegraphics[width=0.95\linewidth]{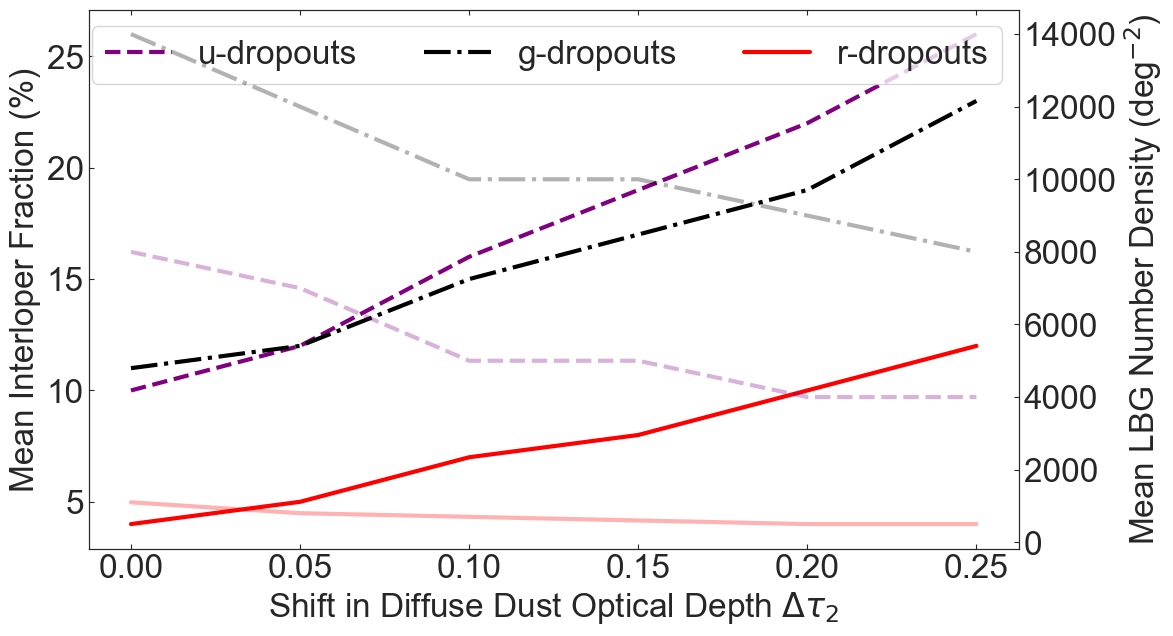}
    \caption{The expected interloper fraction (bold curves) and detected LBG number density (faint curves) produced by perturbing the \textsc{popcosmos} diffuse dust optical depth by $\Delta \tau_{2}$.}
    \label{fig:int_vs_tau}
\end{figure}

To investigate the effect of different levels of ISM dust on the interloper fractions of modelled LBG redshift distributions, we perturb the diffuse optical depth of the \citet{popcosmos} samples as seen in Figure~\ref{fig:int_vs_tau}. We find that perturbing $\tau_{2}$ of the \citet{popcosmos} samples by a constant amount $\Delta \tau_{2}$, increases the fraction of interlopers selected by the LBG selection cuts. This also obscures light from LBGs, leading to fewer detected high-redshift galaxies. If the ISM in LBGs detected by LSST is typically more opaque than predicted by galaxies in \citet{popcosmos}, this could yield less informative constraints on cosmological parameters, due to a stronger mixing of probed scales caused by the larger fraction of interlopers being selected.

\begin{figure}
    \centering
    \includegraphics[width=0.95\linewidth]{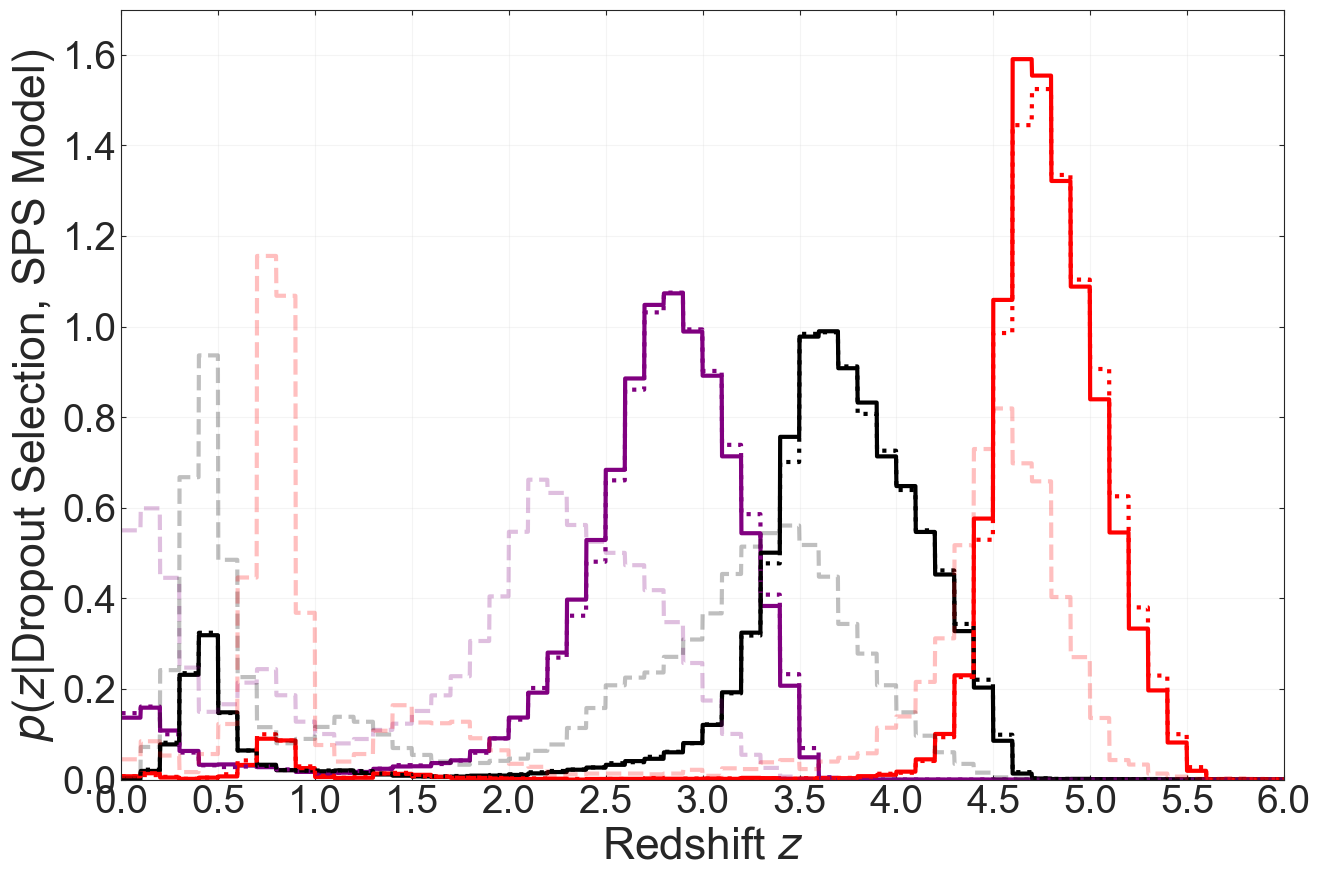}
    \caption{Simulated redshift distributions evaluated at the mean population prior, for a sample of $8\times10^{6}$ galaxies, shown for three different dust models. The distributions implied by \citet{popcosmos}, \citet{nagaraj2022} and Thorp et al. (in prep.) are shown by the solid, dashed, and faint dashed-dotted lines respectively. The u ($z \sim 3$), g ($z \sim 4$) and r ($z \sim 5$) dropouts are denoted in purple, black and red respectively.}
    \label{fig:nzcomp}
\end{figure}

The choice of dust model also has a strong effect on the shape of the redshift distributions as we can see in Figure \ref{fig:nzcomp}. The \citet{nagaraj2022} dust model results not only in just more interlopers, but results in LBG redshift distributions peaking at lower redshifts. This highlights how increased dust in the galaxy population could make it more difficult to select LBGs, especially the higher redshift galaxies in each of the u, g, and r dropout subpopulations. The change in the shape of the distributions due to uncertainty in the population dust model could also have implications on the precision cosmological parameters forecasted in Section \ref{sec:fisher}.

\begin{figure}
    \centering
    \includegraphics[width=0.95\linewidth]{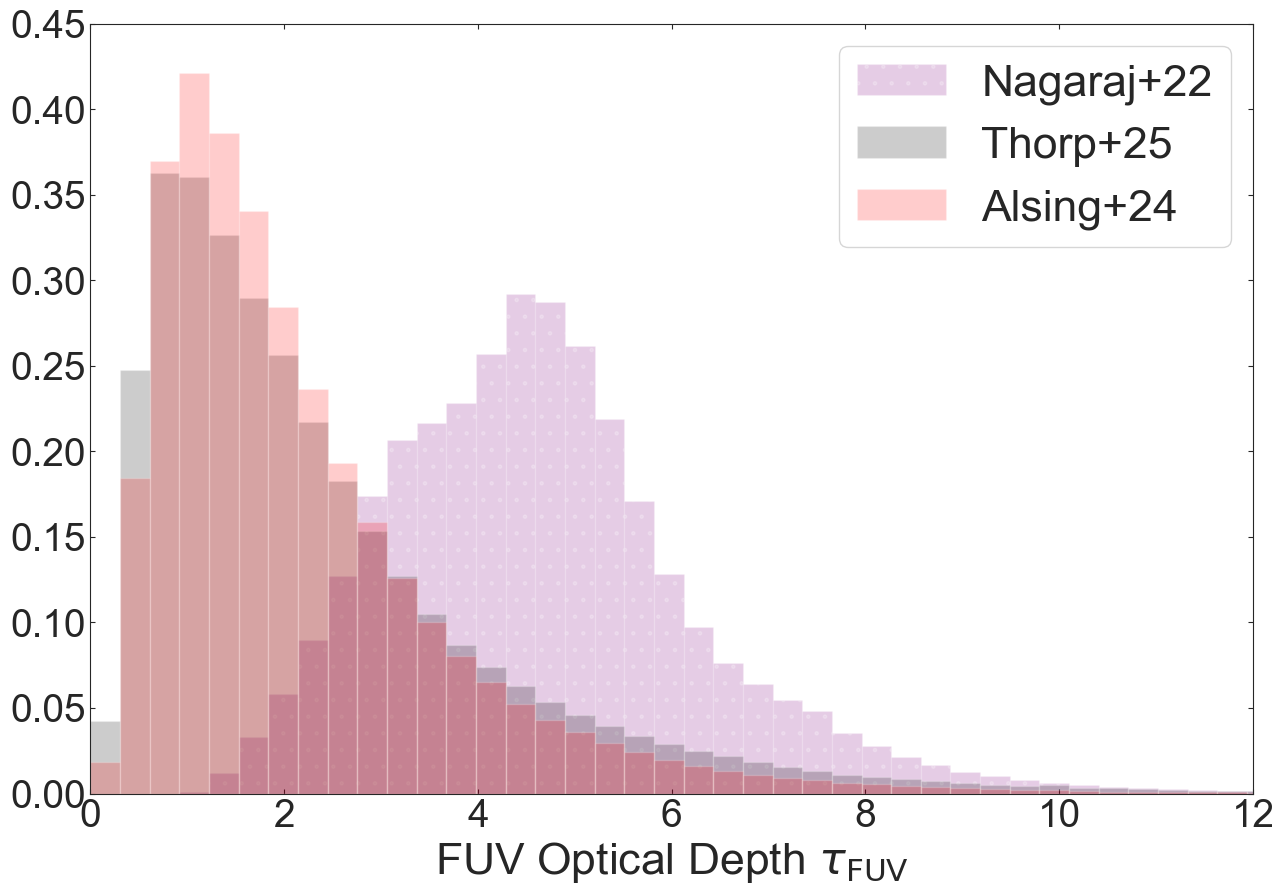}
    \caption{Normalised marginal probability density distribution of the optical depth $\tau_{\mathrm{FUV}}$ implied by \citet{popcosmos},  \citet{nagaraj2022} and Thorp et al. (in prep.). }
    \label{fig:dustmarg}
\end{figure}

\begin{figure*}
    \centering
    \includegraphics[width=0.95\linewidth]{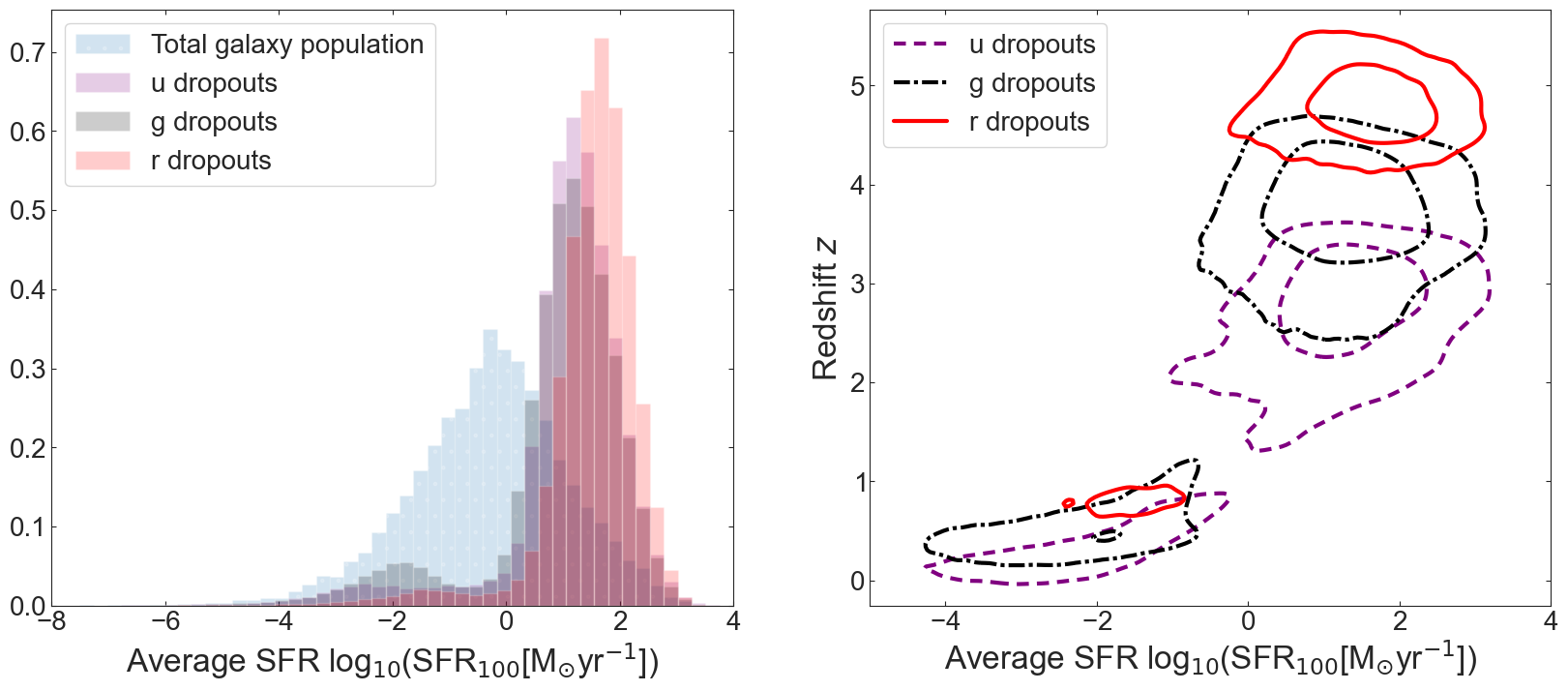}
    \caption{The marginal distributions of galaxies in $\mathrm{log}_{10}\mathrm{SFR}_{100}$ (left panel) that pass the LBG selection defined in Section~\ref{sec:selec}. This is compared with the total population given by the \citet{popcosmos} prior, which is shown by the blue shaded region, showing how selection targets star forming galaxies. The long tails at $\mathrm{log}_{10}\mathrm{SFR}_{100} < 0$ are the lower redshift (quiescent) interlopers, where the redshift dependence can be seen on the right panel, evaluated at the mean of our population prior. The contours represent the 68\% and 95\% confidence limits.}
    \label{fig:alsing24_contour}
\end{figure*}

However, it might be likely that the \citet{nagaraj2022} dust model may be too dusty for the purposes of LBG population modelling between $z \sim $ 3--5. While determining an appropriate dust model for LSST LBGs is beyond the scope of this work, the literature provides some information on the level of ISM dust to expect for LBGs at these redshifts. We find that LBGs at $z \sim $ 3--5 are found to typically to be inside the range of $\tau_{\mathrm{FUV}} \sim $ 1--5 \citep{alvarez16, alvarez19, koprowski18}. We can see clearly in Figure \ref{fig:dustmarg} that the dust model in \citet{nagaraj2022} implies that a significant fraction of galaxies will have $\tau_{\mathrm{FUV}} > 5$. Therefore the dust model implied by \citet{popcosmos} may be more accurate at describing the population distribution of dust in LBGs rather than \citet{nagaraj2022}. However to make this argument, LBGs would have to make up a majority of galaxies at $z \sim $ 3--5. This is supported by measurements of the GSMF at these redshifts \citep{weaver2023, weibel24}, where we see that it is dominated by bright, star forming galaxies like LBGs. We show how LBG selection targets star forming galaxies in our sample in Figure \ref{fig:alsing24_contour}.

However, without knowing the true population distribution of dust of LBGs, it may be likely that the redshift distributions presented in this work (Figure \ref{fig:simnz}) may not be accurately predicting the fraction of interlopers being selected. Even small amounts of extra dust can impact the interloper fraction as was shown in Figure~\ref{fig:int_vs_tau}. However, we see in the study by \citet{koprowski18} that the majority of LBGs selected for their analysis have an ultraviolet continuum slope $\beta < -1.0$. This could imply that the majority of LBGs may have $\tau_{\mathrm{FUV}}$ on the lower dust end of the range $\tau_{\mathrm{FUV}} \sim $ 1--5, similar to the behaviour implied by the \citet{popcosmos} model in Figure \ref{fig:dustmarg}. 

\subsection{Implications for galaxy bias modelling}

\label{sec:intbias}
The bimodal nature of the LBG redshift distributions could have implications for galaxy bias modelling, due to the mixing of scales probed at different redshifts. Concerns about the evolution of galaxy bias across broad redshift distributions has been investigated in the context of weak lensing surveys in \citet{pandey2025}. The work in \citet{schmittfull2018} shows the effect of catastrophic outliers in LBG redshift distributions on cosmological constraints. We investigate the contribution of the interlopers to the clustering signal for LBGs, with a similar approach to \citet{schmittfull2018}, where we instead parametrise the redshift distribution as a Gaussian mixture. We consider a model of a g-dropout redshift distribution, with two Gaussians centred at $z=4$ and $z=0.5$, with standard deviations $0.5$ and $0.1$, for the LBGs and interlopers respectively. Each Gaussian is normalised such that the area under each is given by $f_{\mathrm{int}}$ for the interlopers, and $1-f_{\mathrm{int}}$ for the LBGs. We can assume the populations (LBGs and interlopers) should not overlap (as shown in Figure \ref{fig:simnz}), which means we can simplify the calculation of the full (auto) angular power spectrum $C_{\ell}^{\mathrm{full}}$ by splitting it into two components: an LBG contribution $C_{\ell}^{\mathrm{lbg}}$ and an interloper contribution  $C_{\ell}^{\mathrm{int}}$ such that 
\begin{equation}
    C^{\mathrm{full}}_{\ell} \simeq f_{\mathrm{int}}^{2}C^{\mathrm{int}}_{\ell} + (1-f_{\mathrm{int}})^{2}C^{\mathrm{lbg}}_{\ell}.
\label{eq:twocompangpower}
\end{equation} 

\begin{figure}
    \centering
    \includegraphics[width=0.95\linewidth]{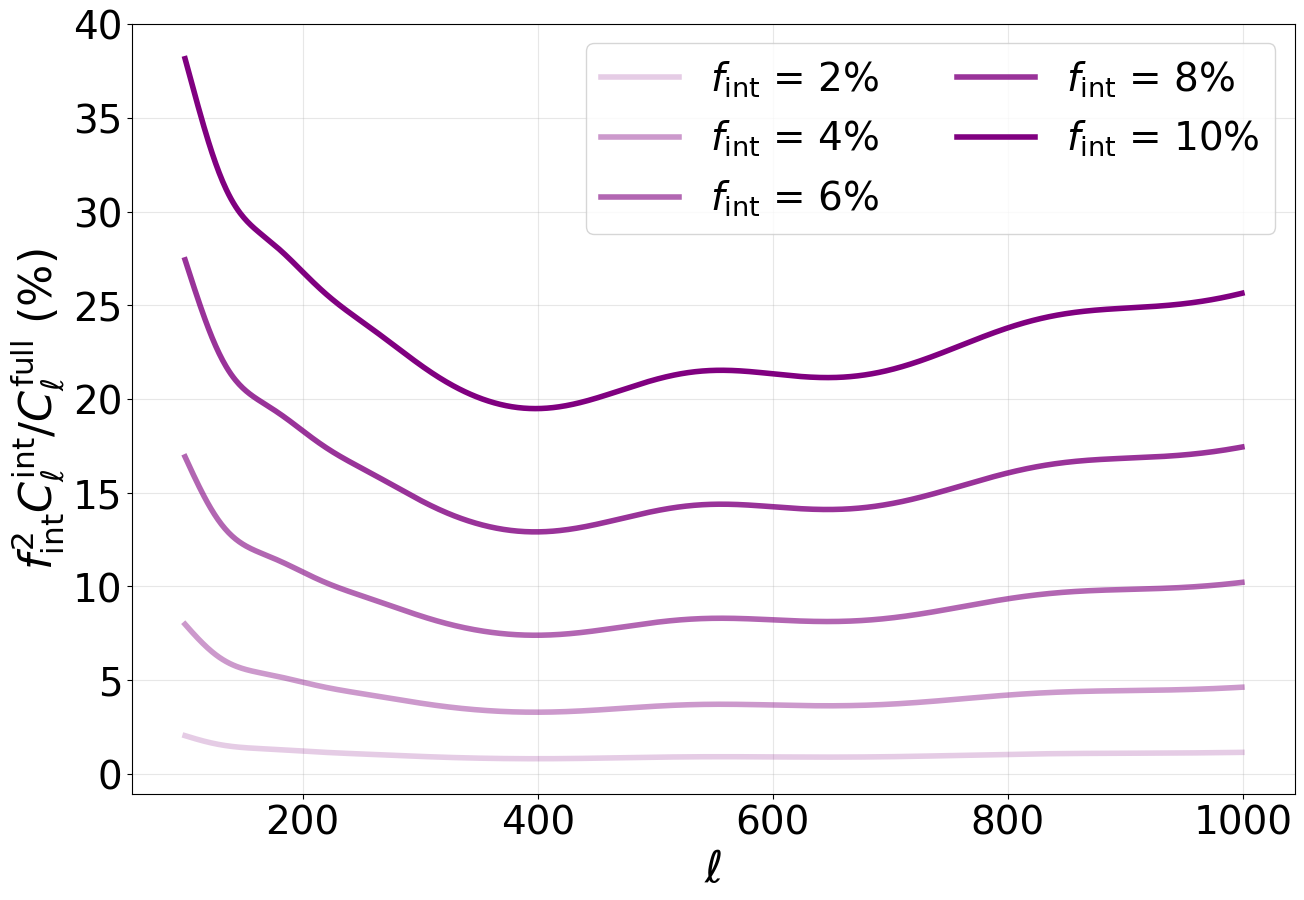}
    \caption{Interloper contribution to the power spectrum shown for different interloper fractions $f_{\mathrm{int}}$ (defined in the text is Section~\ref{sec:nzs}).}
    \label{fig:intbias}
\end{figure}

By calculating the noiseless angular power spectrum for different $f_{\mathrm{int}}$, we can use equation \ref{eq:twocompangpower} to see how the contribution of the interlopers to the total signal $C_{\ell}^{\mathrm{full}}$ depends on $f_{\mathrm{int}}$, as shown in Figure~\ref{fig:intbias}. We can see that the interloper contribution to the total signal increases sharply at larger scales as ($\ell \lesssim 200$), which motivates the scale cut used in Section~\ref{sec:fisher} for our forecast. If $f_{\mathrm{int}}$ becomes too high, the increased mixing of probed scales could mean that a linear bias model may no longer be sufficient for describing the LBG galaxy bias. A more realistic bias model could also allow a greater range of scales to be included in the analysis, compared to our more conservative cuts of $200 \leq \ell \leq 1000$. The Limber approximation could also be dropped to include more of the larger scales, provided $f_{\mathrm{int}}$ is not too large. If the interloper contamination is high, it could be reduced by optimising LBG selection, which is being investigated by work such as \citet{ruhlmann24, payerne25, crenshaw2025}.
\subsection{Limitations of the SPS model}
\label{sec:spslimitations}

In this work we assumed a \citet{madau1995} IGM model, as this is the only model implemented in the current version of FSPS. However, more recent, updated IGM models exist, such as \citet{inoue2014}, which are more accurate at $z\gtrsim5$. This may have an effect on the higher end of the redshift range we are interested in of $3 \lesssim z \lesssim 5$, particularly for the $r$ dropouts at $z\sim 5$. At these redshifts the IGM becomes increasingly opaque, where the Lyman-$\alpha$ damping wing \citep{miralda1998} could become an important factor in LBG selection, which is not included in the SPS model for this work. We also neglect any variation in the IGM optical depth across sight lines.

A further consideration is the modelling of the Lyman-$\alpha$ emission in FSPS. This only includes stellar and nebular emission contributions, and does not include any effects from orientation, inflows, outflows, etc., which can affect the profile of the Lyman-$\alpha$ emission \citep{lyalpha}. Importantly, the resonant scattering of the Lyman-$\alpha$ photons makes the strength of the emission difficult to model \citep{lyalpha}. As LBG dropout selection can be sensitive to the Lyman-$\alpha$ emission, modelling these galaxies incorrectly could affect the shape of the redshift distributions and the expected number densities. This could become even more significant if LBG selection is tailored to select a higher fraction of Lyman-$\alpha$ emitters as experimented in \citet{ruhlmann24}. 

Also, it should be noted our simple factorisation of the population model defined in Equation~\ref{eq:popmodelfac} neglects a number of higher dimensional correlations between SPS parameters. These are present in more sophisticated models like \citet{popcosmos}, where the hope is that this extra information can lead to a more accurate determination of the galaxy population. Therefore we expect to include these correlations in future work as models such as \citet{popcosmos} and Thorp et al. (in prep.) are developed for higher redshift galaxies.

\section{Conclusions}
\label{sec:conclusions}

By incorporating uncertainties in the galaxy population model, we can use SPS modelling to not only forward model LBG redshift distributions, but also provide an estimate of the uncertainty on these distributions. Marginalising over these allows us to provide a forecast on cosmological parameters for LSSTY10 LBGs, assuming no spectroscopic data.

We capture known uncertainties in the GSMF and the CSFRD using Gaussian processes, yielding a continuous parametrisation of the redshift evolution of the GSMF, including its uncertainties between 0 $< z \leq 7$. Using these models, we can generate different realisations of galaxy population distribution in redshift, logarithmic mass and SFH consistent with observations. For other physical parameters, which we find do not have a large effect on the redshift distributions, we assume a much more flexible Gaussian prior. While this likely overestimates the uncertainty from these sources, this approach is taken for the purposes of a more conservative Fisher forecast. 

The resulting redshift distributions generated from samples of different realisations of the galaxy population are qualitatively consistent with LBG distributions from past surveys, and we predict number densities for LSSTY10 in agreement with previous estimates. Due to the flexibility introduced into the galaxy population model, we have shown a method for using SPS to not only forward model redshift distributions, but also associated uncertainties stemming from the galaxy population model, building upon the work in \citet{alsing2023}. This has enabled us to begin to put an estimate on the uncertainty on the expected number densities for LSSTY10 LBGs, improving upon the estimates in \citet{wilson19}.

In addition, this flexibility introduced in the population model allows us to build a redshift distribution prior, which can be marginalised over for a cosmological analysis. This has made it possible to produce an LSSTY10 forecast on cosmological parameters, using SPS simulated LBG redshift distributions. We have shown that for a \citet{popcosmos} dust model,  photometric data alone can provide constraints on $\sigma_{8}$ at a similar precision to those inferred from {\em Planck} observations of the CMB anisotropies \citep{planck18}. This could potentially provide insight into the $S_{8}$ `tension' independently of other data, to probe the matter density before dark energy-dominated times. 

We find that our results are sensitive to the choice of dust attenuation model, which we have shown to significantly affect the number of low redshift contaminants in the LBG redshift distributions. This could lead to less informative constraints on cosmological parameters, as predicted from our fiducial population prior assuming an \citet{popcosmos} dust model. However we argue that if we compare the $\tau_{\mathrm{FUV}}$ values for each dust attenuation model to observations in the literature, we estimate that either an \citet{popcosmos} or Thorp et al. (in prep.) dust model may more accurately characterise the dust in LBGs than \citet{nagaraj2022}. Given the potential of limited availability of spectroscopic LBG data at high redshift, this shows how accurately determining the population dust model of LBGs will be crucial for exploiting these galaxies for cosmological parameter inference. Currently, galaxy population models such as \citet{popcosmos} are limited to redshifts below the range targeted by this work of $3 \lesssim z \lesssim 5$, however newer iterations such as in Thorp et al. (in prep.), are beginning to extend deeper into this regime. With the help of surveys such as JWST, LSST, and Euclid, we will be able to calibrate future SPS and population models to higher and higher redshifts. This will hopefully facilitate the development of more advanced galaxy population models, that will be required for more accurate $3 \lesssim z \lesssim 5$ LBG redshift distribution modelling.

\section*{Acknowledgements}

BL is supported by the Royal Society through a University Research Fellowship.
ST, JA, HVP, and SD have been supported by funding from the European Research Council (ERC) under the European Union's Horizon 2020 research and innovation programmes (grant agreement no.\ 101018897 CosmicExplorer). 

%%%%%%%%%%%%%%%%%%%%%%%%%%%%%%%%%%%%%%%%%%%%%%%%%%
%\section*{Data Availability}



\bibliographystyle{mnras}
\bibliography{main_doc} % if your bibtex file is called example.bib

%\begin{thebibliography}{99}
%\end{thebibliography}

%%%%%%%%%%%%%%%%%%%%%%%%%%%%%%%%%%%%%%%%%%%%%%%%%%

%%%%%%%%%%%%%%%%% APPENDICES %%%%%%%%%%%%%%%%%%%%%

%%%%%%%%%%%%%%%%%%%%%%%%%%%%%%%%%%%%%%%%%%%%%%%%%%

% Don't change these lines
\bsp	% typesetting comment
\label{lastpage}
\end{document}